%% file: main.tex
\documentclass[sn-nature]{sn-jnl}

\usepackage{amsmath}
\usepackage{graphicx}%
\usepackage{multirow}%
\usepackage{amssymb,amsfonts}%
\usepackage{amsthm}%
\usepackage{mathrsfs}%
\usepackage[title]{appendix}%
\usepackage{xcolor}%
\usepackage{textcomp}%
\usepackage{manyfoot}%
\usepackage{booktabs}%
\usepackage{algorithm}%
\usepackage{algorithmicx}%
\usepackage{algpseudocode}%
\usepackage{listings}%
\usepackage{bm} 
\usepackage{upgreek}
\usepackage{subfiles}
\usepackage{lineno}
\usepackage{siunitx}
\usepackage{cases}
\usepackage{microtype}
\usepackage{amssymb}
\usepackage{siunitx}
\usepackage{subfiles}
\usepackage{float}
\usepackage{dblfloatfix} 
\usepackage{nameref}
\usepackage{tabularx}
\usepackage{longtable} 
\usepackage{colortbl}
\usepackage{placeins}
\usepackage{lipsum}

\renewcommand{\figurename}{Figure}
\newcommand\numberthis{\addtocounter{equation}{1}\tag{\theequation}}
\newcommand{\mat}[1]{{\boldsymbol{\mathbf{#1}}}}
\renewcommand{\vec}[1]{{\boldsymbol{\mathbf{#1}}}}

\newcommand{\diagop}{{\text{diag}}}

\renewcommand{\Re}[1]{{\text{Re}(#1)}}
\renewcommand{\Im}[1]{{\text{Im}(#1)}}
\newcommand{\tnorm}[1]{{\left\vert\kern-0.25ex\left\vert #1 
    \right\vert\kern-0.25ex\right\vert_2}}
\newcommand{\tnormsq}[1]{{\left\vert\kern-0.25ex\left\vert #1 
    \right\vert\kern-0.25ex\right\vert_2^2}}
\newcommand{\inorm}[1]{{\left\vert\kern-0.25ex\left\vert #1 
    \right\vert\kern-0.25ex\right\vert_\infty}}
\newcommand{\norm}[1]{{\left\vert\kern-0.25ex\left\vert #1 
    \right\vert\kern-0.25ex\right\vert}}
\newcommand{\hl}[1]{{\textcolor{black}{#1}}}

\DeclareMathOperator*{\argmin}{arg\,min}

\begin{document}
\title[Article Title]{Continuous-time nonlinear closed-loop in-memory computing for high-accuracy massive MIMO detection}

\author*[1†]{\fnm{Piergiulio} \sur{Mannocci}}\email{piergiulio.mannocci@polimi.it}
\author[2]{\fnm{Giacomo} \sur{Pedretti}}
\author[1]{\hl{\fnm{Giacomo} \sur{Larelli}}}
\author[3]{\fnm{Fabian} \sur{B\"ohm}}
\author*[3]{\fnm{Thomas} \sur{Van Vaerenbergh}}\email{thomas.van-vaerenbergh@hpe.com}

\affil[1]{\orgdiv{Dipartimento di Elettronica, Informazione e Bioingegneria}, \orgname{Politecnico di Milano}, \orgaddress{\street{Piazza Leonardo da Vinci 32}, \city{Milano}, \postcode{20133}, \country{Italy}}}

\affil[2]{\orgdiv{Large Scale Integrated Photonics (LSIP)}, \orgname{Hewlett Packard Labs}, \orgaddress{\city{Milpitas}, \state{CA}, \country{USA}}}

\affil[3]{\orgdiv{Large Scale Integrated Photonics (LSIP)}, \orgname{Hewlett Packard Labs}, \orgaddress{\city{Brussels}, \country{Belgium}}}

\affil[†]{Present address: \orgname{Unconventional AI}, \orgaddress{\city{Mountain View}, \state{CA}, \country{USA}}}

\abstract{Analog in-memory computing (IMC) has emerged as a promising approach for accelerating matrix operations by exploiting the intrinsic physics of memory arrays. To date, however, most IMC architectures have focused on linear algebra workloads in which computation is encoded in the equilibrium state of a physical system. Extending these principles to nonlinear optimization remains challenging and typically requires iterative algorithms composed of repeated linear operations.

Here, we introduce a continuous-time nonlinear closed-loop IMC architecture for box-constrained zero-forcing (BCZF) decoding in massive multiple-input multiple-output (MIMO) systems. The proposed architecture embeds the decoding problem directly within the dynamics of a nonlinear feedback network composed of memory arrays and supply-limited operational amplifiers, allowing solutions to emerge through continuous-time physical optimization. We derive a compact analytical model of the circuit and show that its trajectories correspond to the minimization of an equivalent energy function. Experimental emulation using a fabricated IMC chip confirms the predicted dynamics and energy-minimization behavior under realistic hardware nonidealities on up to 16×16 MIMO systems.

To overcome the finite precision of analog hardware, we further extend mixed-precision iterative refinement from linear algebra operations to nonlinear continuous-time optimization. The resulting hybrid analog–digital architecture combines the efficiency of a low-precision physical solver with high-precision residual correction, enabling reliable detection of high-order modulation formats including \hl{1024-QAM}. Benchmark projections indicate operation across a broad spectrum ranging from ultra-low-energy approximate decoding to high-accuracy massive MIMO detection. Together, these results extend closed-loop IMC from equilibrium-based linear algebra to continuous-time nonlinear optimization and establish a pathway toward efficient physical accelerators for high-accuracy wireless communications.}


\maketitle

\section{Introduction}\label{sec:intro}
\subfile{sections/sec1}

\section{Continuous-time physical optimization for MIMO detection}\label{sec:mimo_intr}
\subfile{sections/sec2}

\section{In-memory BCZF circuit}\label{sec:imc_bczf}
\subfile{sections/sec3}

\section{In-memory BCZF detection accuracy}\label{sec:mimo_results}
\subfile{sections/sec4}

\section{Experimental validation}\label{sec:experimental}
\subfile{sections/sec5}

\section{Nonlinearity-aware iterative refinement}\label{sec:refinement}
\subfile{sections/sec6}

\section{Benchmark and scaling}\label{sec:bench}
\subfile{sections/sec7}

\section{Discussion}\label{sec:concl}
\subfile{sections/sec8}

\section{Methods}
\subfile{sections/methods}

\clearpage\backmatter


\section*{Declarations}

\noindent\textbf{Funding}
This work was supported by the European Research Council (ERC) under the European Union’s Horizon Europe Research and Innovation Programme under Grant ERC-2021-AdG-101054098-ANIMATE, and by the Defense Advanced Research Projects Agency (DARPA) under Air Force Research Laboratory (AFRL) contract no FA8650-23-3-7313.

\noindent\textbf{Competing interests} 
The authors declare no competing interests. 

\noindent\textbf{Ethics approval} 
Not applicable. 

\noindent\textbf{Consent to participate}
Not applicable.

\noindent\textbf{Consent for publication}
Not applicable.

\noindent\textbf{Availability of data and materials}
Experimental data shown in figures in this work are available via Zenodo at https://zenodo.org/records/23462745 (ref.~\cite{mannocci_supporting_2026}). Additional characterization data are available from the corresponding authors upon reasonable request. 

\noindent\textbf{Code availability} 
All program codes used in this work are available from the authors upon reasonable request. 

\noindent\textbf{Authors' contributions}
G.P. and T.V.V. conceived the project and established the research framework. P.M. conceived and designed the IMC-BCZF circuit, developed the analytical model, and performed the MIMO decoding simulations. \hl{P.M. and G.L. carried out the experimental validation.} P.M. and T.V.V. conceived and developed the iterative-refinement and discrete-time emulation frameworks. G.P., T.V.V., and F.B. contributed to the analysis and interpretation of the circuit simulations and MIMO decoding results. T.V.V. contributed to the analysis of the experimental validation results. P.M., T.V.V., and F.B. contributed to the benchmarking methodology and performance analysis. All authors contributed to the discussion of the results and the preparation of the manuscript.

\clearpage
\bibliography{references}

\clearpage
\subfile{supplementary}
\end{document}

%% file: sections/sec1.tex
The growing computational demands of modern information processing systems have renewed interest in physical computing paradigms that exploit the intrinsic dynamics of electronic devices to perform computation directly in hardware~\cite{aguirre_hardware_2024}. Among these approaches, analog in-memory computing (IMC) has emerged as a promising candidate for overcoming the energy and throughput limitations of conventional digital architectures by co-locating memory and computation within the same physical substrate~\cite{ielmini_-memory_2018,singh_design_2025}. Using arrays of programmable resistive memories, IMC systems can execute matrix operations such as matrix-vector multiplication (MVM) directly through Ohm's and Kirchhoff's laws, enabling highly efficient acceleration of workloads ranging from machine learning and scientific computing to signal processing and communications~\cite{mannocci_-memory_2023,lepri_-memory_2023,sebastian_memory_2020}.

To date, the vast majority of IMC systems have focused on accelerating linear algebra operations. In open-loop architectures, matrix computations are performed through the steady-state response of memory arrays coupled with dedicated readout peripherals. More recently, closed-loop architectures have extended this concept by embedding memory arrays within feedback networks, enabling efficient implementation of inverse matrix-vector multiplication (IMVM), matrix inversion, and related linear algebra primitives~\cite{sun_solving_2019,mannocci_generalized_2023,mannocci_fully_2026}. Despite their apparent differences, these approaches share a common computational principle: the problem is encoded in the equilibrium state of a physical system, while the transient dynamics merely determine the time required to reach the solution.

Extending this paradigm to nonlinear optimization remains challenging. Many practically relevant tasks, including constrained optimization, inverse problems, machine learning, and signal detection, cannot be expressed as a single linear transfer function and are therefore solved through iterative numerical algorithms involving repeated sequences of linear operations~\cite{golub_matrix_2013}. While such algorithms can be accelerated using IMC primitives, the resulting architectures remain fundamentally discrete-time systems whose behavior depends on algorithmic parameters such as step size, iteration count, and numerical precision.

\hl{An alternative possibility is to directly embed the optimization problem within the continuous-time dynamics of a physical system~\cite{mannocci_iscas_2025,shan_one-step_2024,wang_-memory_2023}. In this framework, the nonlinear circuit dynamics physically implement the optimization process as the system relaxes toward an equilibrium corresponding to the desired solution, rather than relying on a sequence of explicitly discretized update steps. This principle has recently been explored for problems such as compressed-sensing reconstruction using nonlinear closed-loop circuits with supply-limited operational amplifiers. Despite these promising demonstrations, continuous-time nonlinear optimization remains comparatively unexplored within in-memory computing, particularly for large-scale communication problems and high-accuracy computation.}

Massive multiple-input multiple-output (MIMO) detection provides an attractive application domain in which to investigate this paradigm. Massive MIMO is a key enabling technology for current and future wireless networks, allowing base stations equipped with tens or hundreds of antennas to simultaneously serve multiple users over the same time-frequency resources~\cite{chataut_massive_2020}. Recovering the transmitted symbols requires solving a high-dimensional inverse problem involving the channel matrix and the received signal. Existing hardware accelerators span GPUs, FPGAs, ASICs, and analog IMC architectures. Digital implementations provide high numerical precision but incur substantial energy and latency costs due to repeated matrix operations~\cite{shahabuddin_admm-based_2021,tang_058-mm2_2021,jeon_354_2019}. Existing IMC approaches improve efficiency by accelerating individual linear algebra kernels, although demonstrations have been limited to either low-accuracy linear decoders~\cite{mannocci_analogue_2022,mannocci_accelerating_2023,zuo_precise_2025} or rely on discrete-time iterative algorithms to handle nonlinear optimization~\cite{bi_high-speed_2025}. More recently, alternative formulations based on Ising-machine dynamics have also been explored for MIMO detection~\cite{zhu_fully_2026,hashemi_performance_2026}.

In this work, we introduce a nonlinear closed-loop in-memory computing architecture that directly encodes box-constrained MIMO detection within the continuous-time dynamics of a physical system. Unlike previous closed-loop IMC architectures, which compute through their steady-state transfer functions, the proposed approach exploits the dynamical evolution of a nonlinear feedback network as the computational mechanism itself. We show that the resulting circuit naturally implements box-constrained zero-forcing (BCZF) decoding and that its trajectories can be rigorously interpreted as the minimization of an equivalent energy function. A compact analytical model is developed to describe both the transient and steady-state behavior of the system and is experimentally validated through emulation on a fabricated in-memory computing chip.

Because nonlinear physical computation remains subject to the finite precision of analog hardware, we further augment the proposed architecture with a hybrid analog/digital iterative refinement framework. Building upon previous mixed-precision approaches developed for linear algebra operations~\cite{le_gallo_mixed-precision_2018,zuo_precise_2025}, we extend iterative refinement to nonlinear continuous-time physical optimization. The resulting architecture combines the efficiency of a low-precision nonlinear solver with the accuracy of high-precision residual correction, enabling operation across a broad spectrum ranging from ultra-low-energy approximate decoding to high-accuracy detection of high-order modulation formats.

The contributions of this work are fourfold. First, we extend closed-loop in-memory computing from equilibrium-based linear algebra operations to trajectory-based nonlinear optimization by directly encoding the target problem within the continuous-time dynamics of a physical system. Second, we demonstrate that the resulting nonlinear feedback network naturally implements box-constrained MIMO detection through the minimization of an equivalent energy function. Third, we experimentally validate the underlying computational dynamics using fabricated in-memory computing hardware. Finally, we introduce an iterative refinement framework enabling high-accuracy solutions to be recovered from low-precision physical computation. Together, these results establish continuous-time nonlinear dynamics as a new computational primitive for in-memory computing and suggest a broader pathway toward efficient physical optimization beyond conventional linear algebra workloads.

%% file: sections/sec2.tex
\begin{figure}
\centering
\includegraphics[width=\textwidth]{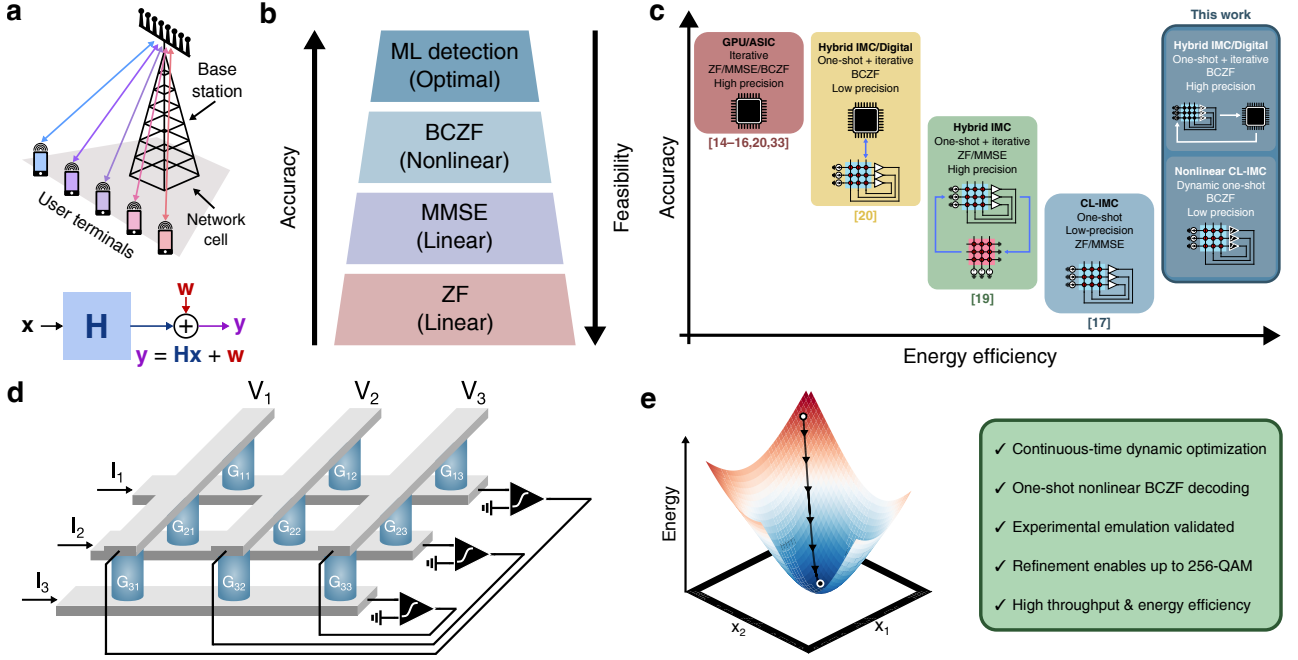}
\caption{\textbf{Overview of continuous-time physical optimization for massive MIMO detection.} (a)~Massive MIMO uplink scenario, where the transmitted symbol vector $\vec{x}$ propagates through the channel matrix $\mat{H}$ and is corrupted by noise $\vec{w}$, producing the received signal $\vec{y}=\mat{H}\vec{x}+\vec{w}$. (b)~Hierarchy of MIMO detection algorithms, illustrating the tradeoff between decoding performance and computational complexity. BCZF approaches ML accuracy while maintaining polynomial complexity. (c)~Positioning of existing MIMO detection accelerators. Digital platforms (GPUs, FPGAs, ASICs) provide high numerical precision but rely on explicit arithmetic operations, whereas IMC architectures exploit physical matrix operations for improved efficiency. The proposed work introduces continuous-time nonlinear physical optimization. (d)~Conceptual architecture of the proposed decoder. The channel matrix is embedded within a nonlinear closed-loop IMC system composed of memory arrays and nonlinear operational amplifiers. (e)~Energy-minimization interpretation of the proposed framework. The circuit dynamics naturally evolve toward minima of an equivalent energy function corresponding to BCZF solutions, extending closed-loop IMC from equilibrium-based linear algebra to trajectory-based nonlinear optimization.}
\label{fig:overview}
\end{figure}

Fig.~\ref{fig:overview} provides an overview of the proposed framework, from the target application domain to the underlying computational paradigm. Fig.~\ref{fig:overview}a illustrates the considered massive MIMO uplink scenario. Each of the $N_t$ users transmits a symbol of energy $E_S$ drawn from a finite constellation $\mathcal{C}$, forming the transmitted vector $\vec{x}\in\mathbb{C}^{N_t}$. A typical modulation scheme is M-ary quadrature amplitude modulation (QAM), where $M$ denotes the QAM depth, \textit{i.e.} the number of symbols in the constellation (\textit{e.g.}, 16-QAM comprises 16 discrete symbols, each carrying 4 information bits). After propagation through the channel matrix $\mat{H}\in\mathbb{C}^{N_r\times N_t}$ and corruption by additive noise $\vec{w}$ ($w_i \sim \mathcal{N}(0,N_0)$), the receiver observes:
\begin{equation}
\vec{y} = \mat{H}\vec{x} + \vec{w}.
\end{equation}
The decoding task consists of recovering $\vec{x}$ from the received signal $\vec{y}$ and the channel state information $\mat{H}$. Several decoding strategies have been proposed, spanning a broad spectrum of accuracy and computational complexity (Fig.~\ref{fig:overview}b). At one extreme, maximum-likelihood (ML) decoding directly searches over all possible constellation vectors,
\begin{equation}
\vec{\hat{x}}_{\mathrm{ML}} = \argmin_{\vec{x}\in\mathcal{C}^{N_t}}\frac12\tnormsq{\mat{H}\vec{x}-\vec{y}},
\end{equation}
yielding optimal detection performance at the expense of exponential complexity~\cite{albreem_massive_2019}. At the opposite extreme, linear detectors such as ZF and MMSE relax the search space to the continuous domain, allowing efficient closed-form solutions but exhibiting degraded performance in heavily loaded systems and for high-order modulation formats, namely: 
\begin{equation}
\vec{\hat{x}}_{\mathrm{L}} = Q\big[\argmin_{\vec{x}}\frac12\tnormsq{\mat{H}\vec{x}-\vec{y}} + \lambda\tnormsq{x}\big],
\end{equation}
where $\lambda = 0$ for ZF and $\lambda = N_0/E_S$ for MMSE, and $Q(\cdot)$ is a suitable \textit{a-posteriori} quantization function matching the selected constellation symbols. Box-constrained zero-forcing (BCZF) occupies an intermediate position by restricting the search to the bounded region containing all admissible constellation points,
\begin{equation}
\vec{\hat{x}}_{\mathrm{BCZF}} = Q\big[\argmin_{\vec{x}\in[-B,B]^{N_t}}\frac12\tnormsq{\mat{H}\vec{x}-\vec{y}}\big],
\label{eq:bczf}
\end{equation}
thereby approaching ML accuracy while maintaining polynomial complexity~\cite{thrampoulidis_symbol_2018}. 

The hardware realization of such nonlinear optimization problems remains challenging. As summarized in Fig.~\ref{fig:overview}c, existing solutions span a continuum ranging from fully digital platforms such as GPUs and FPGAs to analog in-memory computing architectures. Digital accelerators provide high numerical accuracy and algorithmic flexibility, but require the explicit execution of large numbers of arithmetic operations~\cite{shahabuddin_admm-based_2021,jeon_354_2019,tang_058-mm2_2021}. Analog IMC systems exploit the physical laws governing memory arrays to efficiently implement matrix operations, substantially reducing data movement and arithmetic cost~\cite{mannocci_analogue_2022,zuo_precise_2025,mannocci_accelerating_2023}. However, both digital and analog implementations of BCZF typically rely on discrete-time optimization algorithms, such as ADMM or projected-gradient methods, which solve the problem through repeated sequences of linear operations~\cite{bi_high-speed_2025}.

The proposed approach follows a fundamentally different strategy. Rather than accelerating a discretized optimization algorithm, we directly encode the target optimization problem within the continuous-time dynamics of a nonlinear physical system. As illustrated in Fig.~\ref{fig:overview}d, the decoder consists of a nonlinear feedback network combining resistive memory arrays and nonlinear operational amplifiers. The channel matrix is embedded within the feedback paths of the circuit, while amplifier saturation introduces the nonlinear constraints required by BCZF~\cite{mannocci_iscas_2025}. Computation thus emerges from the natural evolution of the physical system toward equilibrium rather than from the sequential execution of arithmetic instructions. The transient trajectory can be interpreted as the minimization of an equivalent energy function whose stationary points correspond to BCZF solutions, as schematically illustrated in Fig.~\ref{fig:overview}e. Unlike previous closed-loop IMC architectures, where dynamics merely determine convergence time, the proposed architecture exploits the dynamics themselves as the computational primitive, extending closed-loop IMC from equilibrium-based linear algebra to trajectory-based nonlinear optimization.

%% file: sections/sec3.tex
\begin{figure}
\centering
\includegraphics[width=\textwidth]{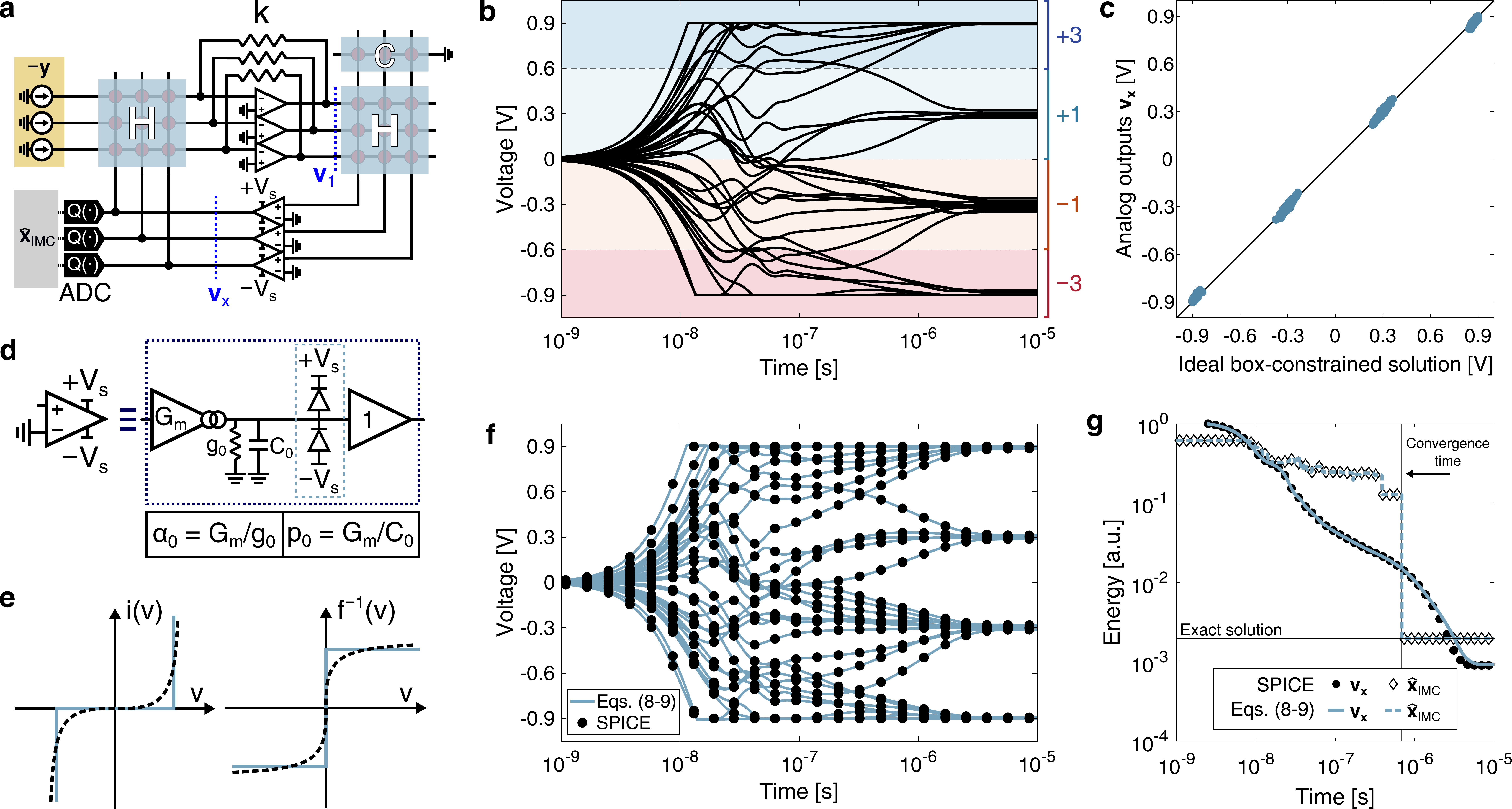}
\caption{\textbf{One-shot IMC-BCZF solver.} (a)~Circuit schematic, composed of two memory arrays enclosed in a feedback loop by TIAs and supply-limited OAs, current DACs injecting input currents, and voltage ADCs for output quantization and readout. (b)~Circuit transient examples for a 16-QAM 16×16 massive MIMO system in low channel noise conditions. Analog voltages $\vec{v}_x$ converge close to exact constellation values. (c)~Analog \hl{voltages $\vec{v}_x$} computed by SPICE as a function of predictions by ideal floating-point 64-bit \hl{box-constrained solution} for a 16×16 16-QAM MIMO system. (d)~Supply-limited OpAmp model. The equivalent transconductance $G_m$, output conductance $g_0$, and capacitance $C_0$ model the linear gain and transient response. $i(v)$ models nonlinear saturation to $\pm V_s$. (e)~Ideal characteristics of the nonlinear network $i(v)$ and inverse discrete map $f^{-1}(v)$ (solid lines) and corresponding smooth approximations used for numerical integration (dashed). (f)~Transient example of IMC-BCZF for a 16-QAM 16×16 MIMO system, as computed by SPICE simulation and by numerical integration of Eqs.~(\ref{eq:ode_1}-\ref{eq:ode_2}). (g)~Equivalent energy $E(\vec{x})$ transient for (f), for the continuous-valued $\vec{v}_x$ state (black dots, continuous line) and post-ADC quantized state $\vec{\hat{x}}_{\text{IMC}}$ (white dots, dashed line), following Eq.~\eqref{eq:bczf_optimality}.}
\label{fig:circuit}
\end{figure}

Fig.~\ref{fig:circuit}a shows the IMC circuit implementing the one-shot solution of Eq.~\eqref{eq:bczf}. Two resistive memory arrays store real-valued replicas \hl{$\mat{H}_R$} of the channel matrix $\mat{H}$~\cite{mannocci_analogue_2022} (see Methods, \nameref{meth:ctor}), with representation of positive and negative matrix values realizable either by differential~\cite{sun_solving_2019} or reference-topology configurations~\cite{mannocci_sram-based_2023}. \hl{For compactness, we denote this real-valued representation simply as $\mat{H}$ in the circuit equations below.} Memory arrays are embedded in a feedback loop by upper and lower OpAmps operating in negative and positive feedback, respectively. A conductance vector $\vec{c}$ is connected between ground and the lower OpAmp inputs to equalize the total load, with \hl{$c_j = \beta - \sum_i|H_{ij}|$ and $\beta = \max\big(\sum_i|H_{ij}|\big)$}. Input currents \hl{representing} $-\vec{y}$ are injected by digital-to-analog converters (DACs) at the virtual grounds of the upper OpAmps, configured as transimpedance amplifiers (TIAs) with local feedback $k$. Analog-to-digital converters (ADCs) enable readout of the decoded estimate $\vec{\hat{x}}_{\text{IMC}}$ by suitably quantizing the outputs $\vec{v}_{\vec{x}}$ of the lower OpAmps, whose supply voltages $\pm V_s$ introduce a saturation nonlinearity in the loop. 

\hl{To relate the numerical and physical domains, we define the constellation-to-voltage scale $V_0=V_s/B$, such that $\vec{v}_{x}=V_0\vec{x}$ and the numerical interval $[-B,B]$ maps onto the physical voltage interval $[-V_s,V_s]$. Correspondingly, the channel and received vector are mapped to conductances and currents as $\mat{G}_\mat{H}=G_0\mat{H}$ and $\vec{i}_\vec{y}=I_0\vec{y}$, respectively, where $I_0=G_0V_0$. It therefore follows that:
\begin{equation*}
    \mat{G}_\mat{H}\vec{v}_{x}-\vec{i}_\vec{y}
    = I_0\left(\mat{H}\vec{x}-\vec{y}\right),
\end{equation*}
so that the physical circuit residual is a scaled replica of the numerical residual in Eq.~\eqref{eq:bczf}. The ADC decision thresholds are scaled by the same voltage factor, equivalently yielding
$\vec{\hat{x}}_{\mathrm{IMC}}
 = Q(\vec{v}_{x}/V_0)$.
For compactness, the multiplicative mapping factors $G_0$, $I_0$, and $V_0$ are omitted in the following circuit equations, and the same symbols $\mat{H}$ and $\vec{y}$ are used for their circuit-mapped counterparts. Post-ADC values thus represent the intended quantized and decoded output, whereas $\vec{v}_{x}$ is the internal continuous-valued proxy.} Notably, ADCs can operate with moderate to low bit-precision in the order of $\log_2(\sqrt{M})$, where $M$ is the QAM modulation depth, \textit{e.g.} $N_{bit} = 2$ for a 16-QAM system. Overall, for an $N\times N$ complex-valued MIMO system, the circuit requires two $2N\times 2N$ real-valued memory arrays, one $2N\times 1$ real-valued memory vector, $4N$ OpAmps, and $2N$ DACs and ADCs.

Fig.~\ref{fig:circuit}b shows a typical transient for a randomly-generated 16×16 MIMO instance, 16-QAM transmitted vector, and low noise power. Notably, voltages \hl{$\vec{v}_\vec{x}$} evolve smoothly over time, with some \hl{OpAmp outputs} exhibiting saturation to supply voltages which, in turn, alter the dynamical trajectory of the circuit. After a transient phase, the \hl{$\vec{v}_\vec{x}$} voltages converge close to \hl{voltage representations $V_0$\{-3, -1, +1, +3\} of the constellation levels}. \hl{Their steady-state condition} can be modeled as the solution of the fixed-point equation: 
\begin{equation}
    \vec{v}_{x} \simeq \text{clip}(-\frac{\alpha_0}{k\beta} \mat{H}^T(\mat{H}\vec{v}_{x} - \vec{y}),-V_s,V_s),
    \label{eq:bczf_circuit_model}
\end{equation}
where $\alpha_0$ is the lower OpAmp open-loop gain. \hl{Here, ‘fixed-point’ refers to the equilibrium of the continuous-time circuit dynamics and does not imply the execution of a numerical fixed-point iteration.} Fig.~\ref{fig:circuit}c compares SPICE-simulated \hl{$\vec{v}_\vec{x}$} voltages with the ideal \hl{pre-quantization} BCZF decoder output, highlighting close matching, with deviations attributable to the finite gain of OpAmps, which in general do not affect the post-ADC output. 

The clipping nonlinearity suggests an interpretation in terms of constrained optimization. Specifically, the fixed-point equation in Eq.~\eqref{eq:bczf_circuit_model} can be shown to correspond to the first-order optimality condition of the equivalent energy function (see Methods, \nameref{meth:energy} for full derivation): 
\begin{align*}
     \vec{v}_{\vec{x}}   &= \argmin_{\vec{v}}E(\vec{v}) \\
                 &= \argmin_{\vec{v}}\bigg( \frac12 \tnormsq{\mat{H}\vec{v}-\vec{y}} + \frac{k\beta}{2\alpha_0}\tnormsq{\vec{v}} + \eta_{[-V_s,V_s]^N}(\vec{v})\bigg) \\
                 &= \argmin_{\vec{v}\in [-V_s,V_s]^N}\bigg( \frac12 \tnormsq{\mat{H}\vec{v}-\vec{y}} + \frac{k\beta}{2\alpha_0}\tnormsq{\vec{v}}\bigg)   \numberthis
 \label{eq:bczf_optimality}
\end{align*}
where $\eta_{[-V_s,V_s]}(x)$ denotes the indicator function of the box constraint (see Methods, \nameref{meth:energy}). For sufficiently large $\alpha_0$, the post-ADC output reads: 
\begin{equation}
    \hat{\vec{x}}_{\text{IMC}} = Q\bigg(\hl{\frac{\vec{v}_{\vec{x}}}{V_0}}\bigg) \simeq Q\bigg[ \argmin_{\vec{x}\in [\hl{-B,B}]^N} \frac12 \tnormsq{\mat{H}\vec{x}-\vec{y}} \bigg] = \hl{\hat{\vec{x}}_{\text{BCZF}}},
\end{equation}
thus matching Eq.~\eqref{eq:bczf} as intended. 

Unlike digital or iterative IMC solvers~\cite{shahabuddin_admm-based_2021,bi_high-speed_2025}, the circuit reaches the solution through a combination of its continuous-time dynamics and the analog saturation nonlinearity of the circuit OpAmps. Fig.~\ref{fig:circuit}d shows the equivalent supply-limited single-pole OpAmp model, consisting of linear transconductance $G_m$, output resistance $r_0$ and capacitance $C_0$, and a nonlinear diode-like network characterized by $i(v)=0$ for $|v| < V_s$, $i(v)\in[0,+\infty)$ for $v=V_s$, and $i(v)\in(-\infty,0]$ for $v=-V_s$ (Fig.~\ref{fig:circuit}e).

The circuit dynamics can be derived based on the methodology described in~\cite{mannocci_generalized_2023} as follows \hl{(see Methods, \nameref{meth:ct_model})}:
\begin{numcases}{}
    \frac{d\vec{v}_1}{dt} \simeq -p_0\mat{U}_1^{-1}(k\vec{v}_1 + \mat{H}\vec{v}_{\vec{x}} - \vec{y}) \label{eq:ode_1}\\
    \frac{d\vec{v}_{\vec{x}}}{dt} \simeq p_0\bigg(\frac1\beta\mat{H}^T\vec{v}_1 - \frac1\alpha_0\vec{v}_{\vec{x}} - f(\vec{v}_{\vec{x}})\bigg) \label{eq:ode_2}
\end{numcases}
where $\mat{U}_1$ is the diagonal matrix of row sums of the left-side array $\mat{H}$ and upper OpAmps feedback conductances $k$~\cite{sun_solving_2019, mannocci_generalized_2023}, $f(v)=i(v)/G_m$, and $p_0$ is the OpAmp gain-bandwidth product (GBWP). At steady state, $\frac{d\vec{v}_1}{dt}=\frac{d\vec{v}_{\vec{x}}}{dt}=0$, yielding: 
\begin{equation}
    \vec{v}_{\vec{x}} = f^{-1}\big(-\frac{1}{k\beta}\mat{H}^T(\mat{H}\vec{v}_{\vec{x}} - \vec{y}) - \frac{1}{\alpha_0}\vec{v}_{\vec{x}}\big),
    \label{eq:inv_map}
\end{equation}
with $f^{-1}(x)\in[-V_s,+V_s]$ for $x=0$, $+V_s$ for $x>0$, and $-V_s$ for $x<0$ (Fig.~\ref{fig:circuit}e), thus matching Eq.~\eqref{eq:bczf_circuit_model}.

Fig.~\ref{fig:circuit}f compares SPICE transients with numerical integration of Eqs.~(\ref{eq:ode_1}–\ref{eq:ode_2}), where hard nonlinearities were replaced by smooth approximants (Figs.~\ref{fig:circuit}e), showing excellent agreement. Correspondingly, Fig.~\ref{fig:circuit}g shows the evolution of the energy function in Eq.~\eqref{eq:bczf_optimality} for both the analog \hl{voltages} $\vec{v}_{\vec{x}}$ and the post-ADC estimate $\hat{\vec{x}}_{\text{IMC}}$. While the analog state evolves continuously, the post-ADC outputs may undergo both descent and ascent in the discrete space. We define the convergence time $T_{conv}$ as the time required for the quantized outputs to stabilize. Since the dynamics scale with the GBWP, $T_{conv} \propto 1/p_0 = t_0$. In the following, we assume $\sisetup{mode=text} p_0 = \SI{100}{\mega\hertz}$, \textit{i.e.} $\sisetup{mode=text} t_0 = \SI{10}{\nano\second}$.

%% file: sections/sec4.tex
\begin{figure}
\centering
\includegraphics[width=\textwidth]{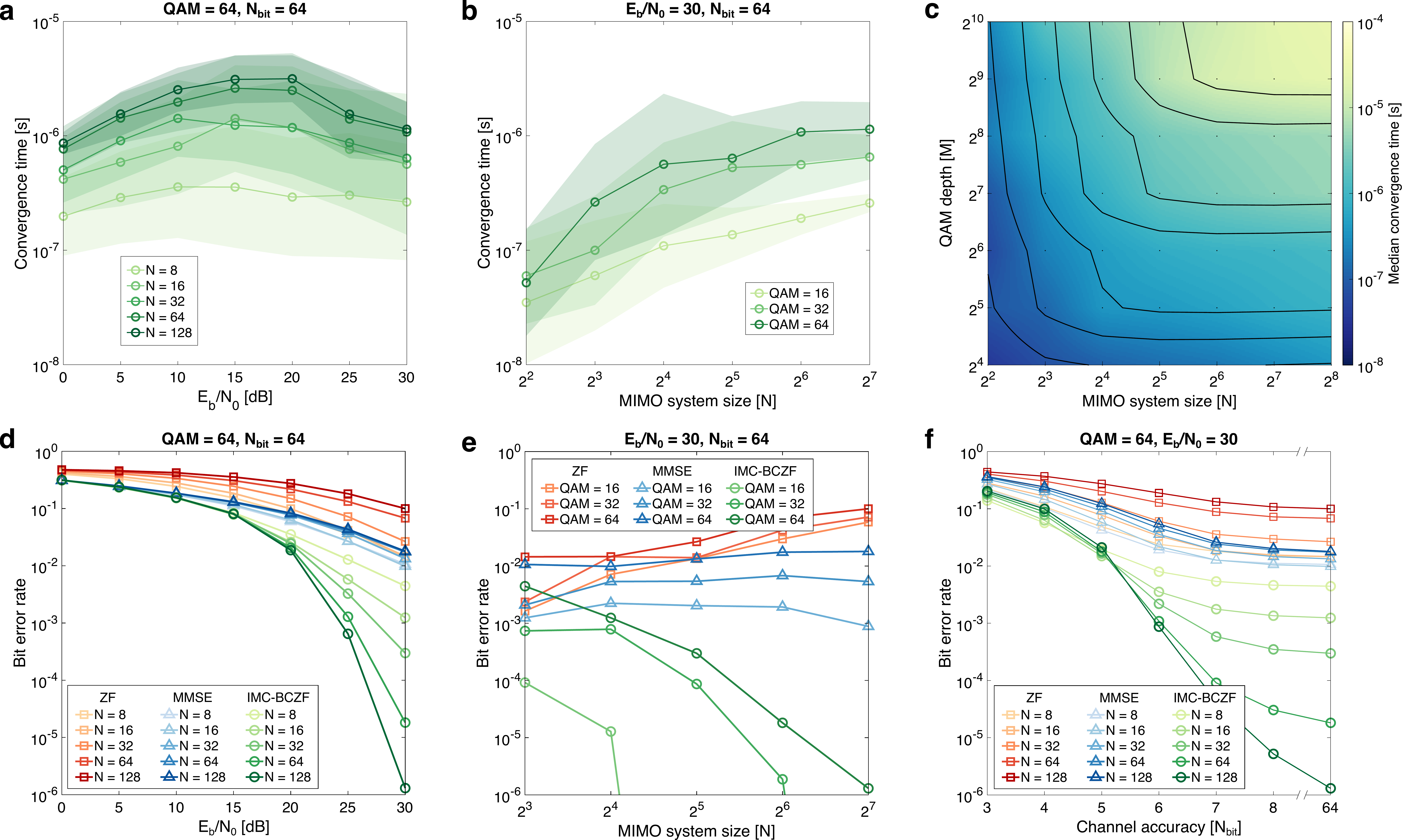}
\caption{\textbf{In-memory BCZF detection performance.} (a,b)~Convergence time as a function of (a)~normalized SNR and increasing MIMO system size at 64-QAM and (b)~MIMO system size and increasing QAM depth at high SNR (line: median, shading: $\pm1\sigma$). \hl{(c)~Median convergence time as a function of MIMO system size and QAM depth, highlighting a progressively decreasing dependence on N as system size increases.} (\hl{d}-\hl{f})~BER as a function of (\hl{d})~normalized SNR, (\hl{e})~MIMO system size and (\hl{f})~bit-accuracy of the memory arrays storing channel matrices $\mat{H}$, increasing system size from $N = 8$ to $N = 128$, and varying QAM modulation depth from $M=16$ to $M=64$. In-memory BCZF consistently outperforms MMSE and ZF, with BER further improving with system size upscaling.} 
\label{fig:mimo}
\end{figure}

To assess the theoretical performance and feasibility of our circuit, we performed extensive simulations of massive MIMO decoding at varying QAM depth $M$, noise power $N_0$, and system size $N$. For each configuration, we averaged over 100 randomly generated Rayleigh-fading MIMO channels, simulating the transmission of at least 10\textsuperscript5 bits per instance (see Methods, \nameref{meth:MIMO_simulation}). 

An important advantage of closed-loop physical computation is that computational latency is determined by the evolution of analog states rather than by the execution of sequential arithmetic operations. Understanding how the convergence time scales with problem parameters is therefore essential for assessing the practicality of the proposed approach. Fig.~\ref{fig:mimo}a reports the simulated convergence time as a function of the normalized SNR $E_b/N_0$ for increasing system sizes ranging from $N = 8$ to $N = 128$, exhibiting only weak dependence on channel noise and remaining nearly constant over the entire investigated SNR range. More importantly, the convergence time remains bounded as the system size increases, as shown also in Fig.~\ref{fig:mimo}b, reporting the convergence time as a function of the system size for varying QAM depths from $M = 16$ to $M = 64$. Even for the largest investigated systems, the growth is substantially slower than the typical $\mathcal{O}(N^{2\sim3})$ complexity of ZF, MMSE, and BCZF in fully-digital solvers~\cite{bi_high-speed_2025}. Unlike the latter, whose complexity is determined by the number of arithmetic operations required to solve increasingly large linear systems, the closed-loop circuit exploits the parallel evolution of all state variables simultaneously~\cite{sun_time_2020,mannocci_generalized_2023}. As a result, computational latency is governed primarily by the physical relaxation dynamics of the network~\cite{mannocci_analogue_2022}. \hl{Furthermore, with respect to fully-linear closed-loop IMC circuits, the saturation nonlinearity further contributes to accelerating the dynamical transient. Whenever one or more OpAmps saturate, the corresponding feedback loops are broken and no longer participate in the circuit dynamics, leaving only a smaller subsystem engaged in the feedback action (see Methods,~\nameref{meth:tts}). The latter typically exhibits smaller associated time-constants, resulting in a saturation-induced speedup which ultimately yields an approximately size-independent $T_{conv} \propto M\log(M)$ for sufficiently large $N$ (Fig.~\ref{fig:mimo}c).} 

We next evaluate the decoding accuracy achieved by the proposed nonlinear solver. Fig.~\ref{fig:mimo}\hl{d} reports the bit error rate (BER) as a function of $E_b/N_0$ for varying system size $N$. For all investigated dimensions, BCZF consistently outperforms the linear ZF and MMSE detectors. Moreover, the relative advantage increases with system size, indicating that the nonlinear box constraint becomes increasingly effective in heavily loaded systems. Higher-order modulation formats represent a more challenging operating regime. Fig.~\ref{fig:mimo}\hl{e} shows the BER as a function of system size for QAM depths ranging from $M=16$ to $M=64$. Although BCZF maintains a clear advantage over linear detection across all investigated modulation formats, decoding accuracy degrades with increasing constellation density due to the reduced distance between neighboring symbols.

Analog memory accuracy constitutes one of the dominant nonidealities in IMC systems~\cite{mannocci_achieving_2026}, with conventional resistive memories typically limited to 5-6 bits of equivalent precision~\cite{pedretti_redundancy_2021,song_programming_2024,pistolesi_drift_2024}. Fig.~\ref{fig:mimo}\hl{f} therefore examines the impact of channel quantization by progressively reducing the precision used to represent the channel matrix, \hl{while assuming the latter to be perfectly known at the receiver}. While ZF and MMSE remain relatively insensitive to quantization over a broad range of precisions, BCZF exhibits a stronger dependence on channel accuracy. Although the resulting performance remains superior to linear detection, the observed sensitivity highlights the importance of accurate channel representations for large-scale nonlinear decoding and motivates the mixed-precision approach introduced in the following sections. \hl{This effect should be distinguished from that of imperfect channel knowledge, for which the detector operates on an estimate
$\hat{\mat{H}}=\mat{H}+\Delta\mat{H}$, where now $\hat{\mat{H}}$ is quantized before being programmed into the array. Consequently, if $\Delta\mat{H}$ does not cause any matrix element to cross a decision boundary of the programming quantizer, then
$Q(\hat{\mat{H}})=Q(\mat{H})$, and the programmed circuit and decoding result remain unchanged. In this regime, the finite programming precision masks the effect of small channel-estimation errors.} 

%% file: sections/sec5.tex
\begin{figure}
\centering
\includegraphics[width=\textwidth]{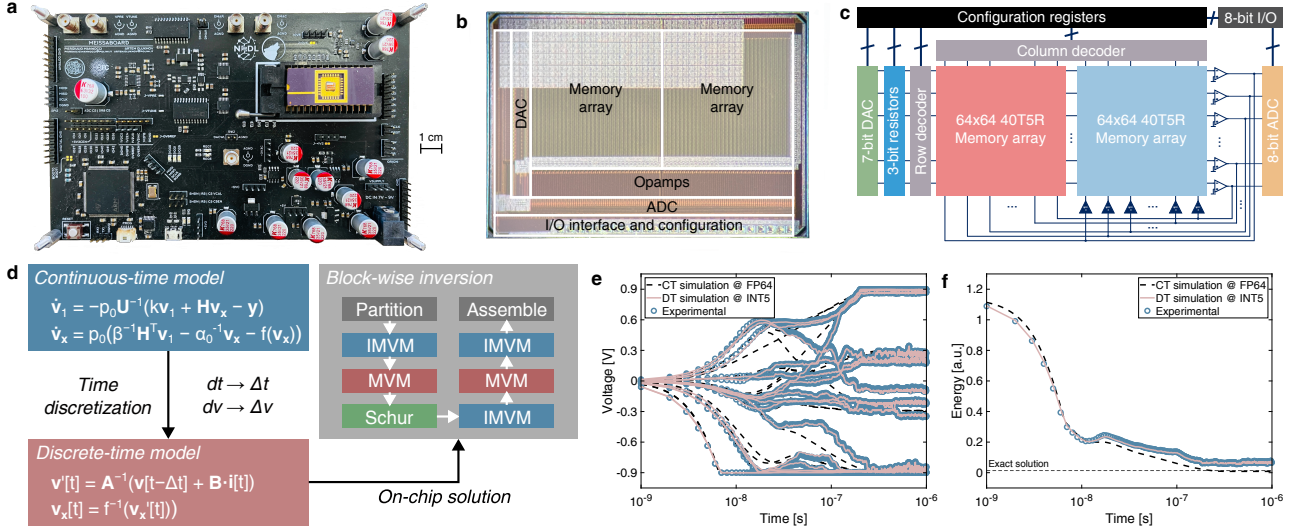}
\caption{\textbf{Experimental emulation of continuous-time dynamics.} (a)~Printed circuit board used for experimental measurements. (b)~Micrograph of the fabricated IMC test chip. (c)~Architecture of the MVM/IMVM-capable testchip used in this work. (d)~Experimental emulation methodology. The continuous-time nonlinear dynamics are discretized and mapped onto a sequence of experimentally measured MVM and IMVM operations executed on the IMC hardware. (e)~Comparison between continuous-time simulation, discrete-time emulation, and experimentally measured trajectories for an 8×8 16-QAM MIMO system, showing close agreement throughout the optimization process. (f)~Evolution of the equivalent energy function for continuous-time simulation, discrete-time emulation, and experimental measurements. The experimentally reconstructed dynamics reproduce the predicted energy-minimization behavior despite hardware nonidealities and limited memory precision.}
\label{fig:experimental}
\end{figure}
To evaluate the robustness of the proposed dynamic circuit to analog nonidealities, we performed an experimental emulation of the closed-loop dynamics described by Eqs.~(\ref{eq:ode_1}-\ref{eq:ode_2}) on a previously-fabricated 90~nm fully-integrated IMC platform designed for acceleration of MVM and IMVM~\cite{mannocci_fully_2026}. Figs.~\ref{fig:experimental}a-c summarize the experimental setup. The hardware platform consists of a custom printed circuit board hosting the mixed-signal IMC linear algebra IC. The chip integrates two $64\times64$ memory arrays together with peripheral DACs, ADCs, transimpedance amplifiers, configuration registers, and digital control logic (see Methods,~\nameref{meth:platform} and Extended Data Fig.~\ref{exfig:chip}).

Notably, direct execution of the continuous-time differential equations introduced in Eqs.~(\ref{eq:ode_1}-\ref{eq:ode_2}) is not possible on the available hardware because the chip was not specifically designed to implement the proposed nonlinear architecture. Instead, we construct a discrete-time approximation of the continuous dynamics and execute each update step using experimentally measured MVM and IMVM operations (see Methods, \nameref{meth:ct_emulation} and Extended Data Fig.~\ref{exfig:ctvsdt}). Fig.~\ref{fig:experimental}e compares the measured transient evolution for an 8×8 16-QAM MIMO system with both full-precision continuous-time simulation and discrete-time numerical integration at the same hardware precision. \hl{Across the expanded experimental campaign, encompassing channel dimensions from 4×4 to 16×16 and three distinct dies, the experimentally measured voltage vectors agree closely with the corresponding ideal trajectories, with a median time-resolved vector relative error of approximately 2\% (see Methods, \nameref{meth:expdataset}).} Moreover, Fig.~\ref{fig:experimental}f reports the evolution of the equivalent energy function introduced in Eq.~\eqref{eq:bczf_optimality} for the three approaches, exhibiting similar energy descent trajectories despite operating under finite precision and device variability. This observation is particularly significant because the energy function is not explicitly programmed into the hardware; rather, it emerges naturally from the interaction of the physical computational primitives. Extended Data Fig.~\ref{exfig:expscale} further confirms this behavior across larger MIMO dimensions, showing that the experimentally emulated trajectories remain in close agreement with the corresponding INT5 discrete-time model. At the same time, the discrepancy between INT5 and full-precision trajectories increases with system size, consistent with the growing sensitivity of BCZF decoding to channel quantization discussed in the previous section. Together, these results validate the physical computational mechanisms underlying the proposed architecture, demonstrating that the continuous-time model accurately predicts hardware behavior and that the corresponding optimization dynamics remain stable under realistic implementation nonidealities. 

%% file: sections/sec6.tex
\begin{figure}
\centering
\includegraphics[width=\textwidth]{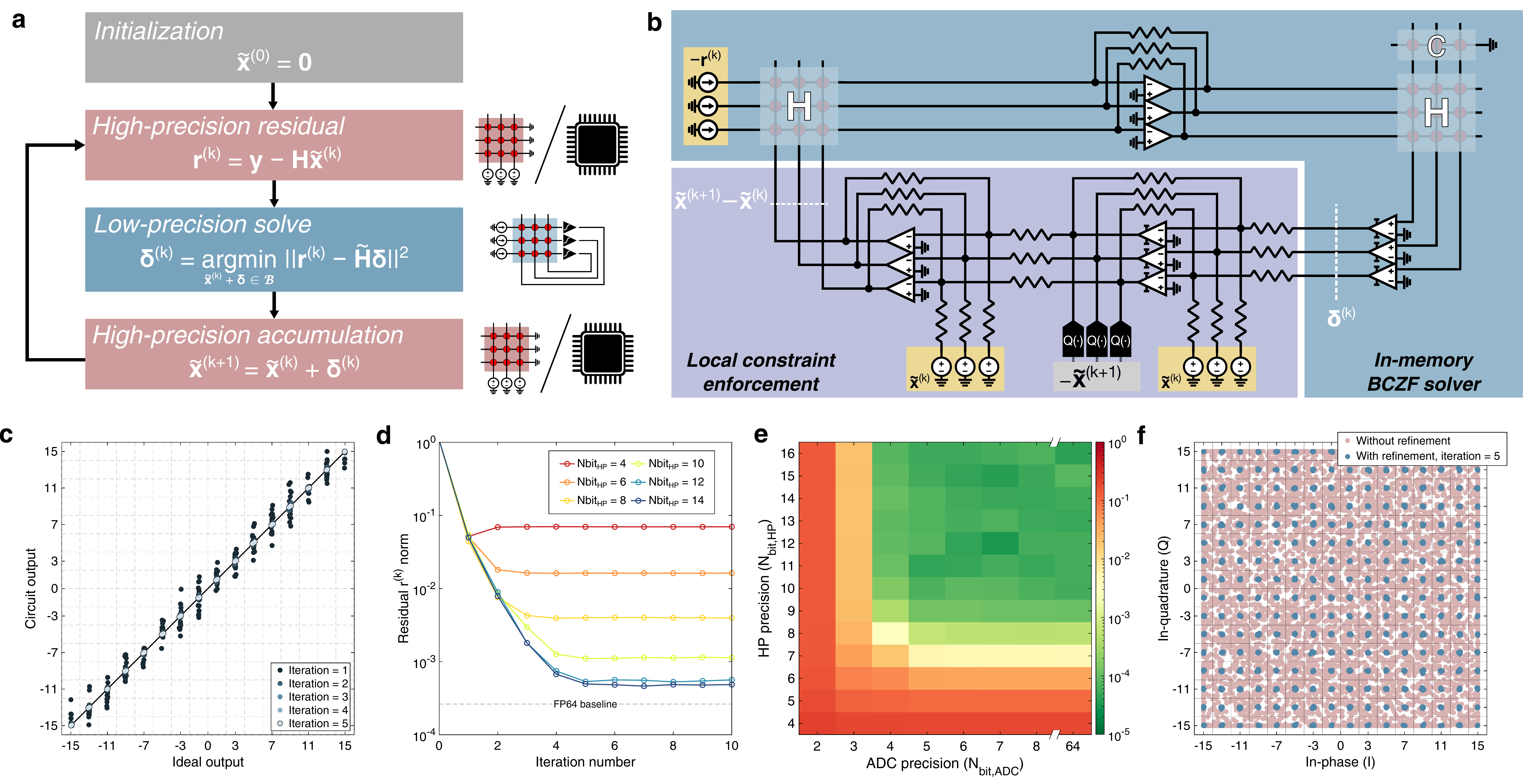}
\caption{\textbf{Iterative refinement for high-precision nonlinear physical optimization.} (a)~Iterative refinement framework. A low-precision nonlinear solver produces an approximate BCZF solution, which is progressively corrected through high-precision residual computation. (b)~Extension of the BCZF circuit of Fig.~\ref{fig:circuit}a through the addition of a local constraint enforcement block within the main feedback loop, operating at a higher speed with respect to the in-memory BCZF solver domain. (c)~Decoded versus exact symbol values for increasing numbers of refinement iterations, illustrating progressive convergence toward the ideal solution. (d)~Solution residual norm as a function of iteration number for different HP-MVM precisions. The ultimate accuracy is limited by the precision of the residual-computation engine. (e)~BER design-space exploration as a function of ADC resolution and HP-MVM precision for a 5-bit memory array with 2\% conductance variability. (f)~Decoded 256-QAM constellation without refinement (red) and after five refinement iterations (blue). Iterative refinement systematically removes the residual error introduced by low-precision physical computation, recovering near-ideal decoding performance.}
\label{fig:ir}
\end{figure}
The finite precision of analog memories and peripheral circuitry ultimately limits the decoding accuracy achievable by the standalone IMC-BCZF solver. Directly increasing analog precision is generally impractical owing to the rapidly increasing cost of high-resolution memories and data converters. Mixed-precision techniques~\cite{le_gallo_mixed-precision_2018,feinberg_analog_2021,zuo_precise_2025} instead combine low-precision computation with high-precision residual correction, enabling high-accuracy solutions at substantially reduced cost. Here, we extend this concept from linear algebra operations to nonlinear physical optimization through the iterative-refinement scheme shown in Fig.~\ref{fig:ir}a. The algorithm begins by initializing the current solution estimate $\tilde{\vec{x}}^{(0)} = \vec{0}$. The residual error at the $k$-th step is then evaluated as:
\begin{equation}
\vec{r}^{(k)} = \vec{y} - \mat{H}\tilde{\vec{x}}^{(k)},
\label{eq:ir_residual}
\end{equation}
where the matrix-vector multiplication is performed using a high-precision (HP) computational unit, either fully-digital~\cite{le_gallo_mixed-precision_2018} or IMC-based~\cite{zuo_precise_2025}. The residual vector is then processed by the same low-precision IMC-BCZF solver introduced previously, producing a correction term:
\begin{equation}
\vec{\delta}^{(k)} \simeq \argmin_{\tilde{\vec{x}}^{(k)} + \vec{\delta} \in \mat{\mathcal{B}}^N} \tnormsq{\vec{r}^{(k)}-\tilde{\mat{H}}\vec{\delta}}
\label{eq:ir_delta}
\end{equation}
where $\tilde{\mat{H}}$ denotes the low-precision replica of the channel matrix $\mat{H}$ mapped in the BCZF circuit. Finally, the correction is accumulated with high precision,
\begin{equation}
\tilde{\vec{x}}^{(k+1)} = \tilde{\vec{x}}^{(k)} + \vec{\delta}^{(k)}.
\label{eq:ir_accumulation}
\end{equation}
The process is repeated until convergence. As in conventional iterative refinement, the low-precision solver captures the dominant error components, while the high-precision backend accumulates increasingly accurate corrections. On the other hand, unlike linear refinement schemes, the proposed nonlinear formulation requires each BCZF solve to be centered around the current estimate $\tilde{\vec{x}}^{(k)}$. This is implemented through the local constraint-enforcement block shown in Fig.~\ref{fig:ir}b, which is embedded within the main feedback loop of the BCZF circuit of Fig.~\ref{fig:circuit}a. Within the constraint-enforcement block, the first voltage-limited OpAmps act as voltage summers between the current estimate $\tilde{\vec{x}}^{(k)}$ and the correction vector $\vec{\delta}^{(k)}$ provided by the in-memory BCZF solver, thus enforcing the corrected estimate to lie within the constellation box as required by Eq.~\eqref{eq:ir_delta}. The sign-inverted sum is fed to the second OpAmps, which similarly operate as voltage summers, feeding $\tilde{\vec{x}}^{(k+1)} - \tilde{\vec{x}}^{(k)}$ back into the analog loop. For stability, OpAmps in the constraint-enforcement block are designed to operate faster than the main BCZF solver.

Fig.~\ref{fig:ir}c visualizes the evolution of the decoded symbols during iterative refinement for a representative 256-QAM detection problem on a 64×64 MIMO channel, encoded into 5-bit analog memories with realistic 2\% variability~\cite{mannocci_fully_2026}. While the initial one-shot solution exhibits noticeable deviations from the ideal constellation points owing to the limited precision of the analog hardware, the symbols progressively converge toward their correct locations as refinement iterations proceed. The convergence process is further quantified in Fig.~\ref{fig:ir}d, which reports the solution residual norm as a function of iteration number for different precisions of the HP residual engine. Similar to linear iterative refinement schemes, the ultimate accuracy achieved by the refinement process is limited solely by the high-precision unit, while the low-precision solver defines the convergence rate of the algorithm~\cite{mannocci_achieving_2026}. 

Fig.~\ref{fig:ir}e explores the impact of ADC/DAC resolution and HP-MVM precision assuming a 5-bit memory array with 2\% variability~\cite{mannocci_fully_2026} (see Extended Data Fig.~\ref{exfig:chip}). The results reveal a clear hierarchy of importance: increasing the precision of the HP residual engine substantially improves the final BER, whereas increasing ADC resolution beyond a modest threshold provides diminishing returns. High detection accuracy can therefore be recovered despite low-precision memories and realistic device variability, demonstrating that accuracy is achieved primarily through algorithmic correction rather than exceptionally precise analog hardware. \hl{In the particular case of imperfect channel knowledge, iterative refinement evaluates the residual using the higher-precision estimate $\hat{\mat{H}}$. Channel-estimation error then introduces the irreducible term $-\Delta\mat{H}\mat{x}$, which acts, to first order, as an additional signal-dependent noise contribution and ultimately limits the effective $E_{\mathrm{b}}/N_0$ and the attainable BER.}

The practical impact of iterative refinement is illustrated in Fig.~\ref{fig:ir}f for a representative 256-QAM detection problem. Without refinement, limited hardware precision causes substantial spreading of the decoded symbols around their ideal locations, resulting in frequent decision errors. After only five refinement iterations, the symbol clusters collapse around the correct constellation points, and the BER approaches that of ideal high-precision decoding. The ability to reliably decode 256-QAM using predominantly low-precision physical computation demonstrates that iterative refinement effectively overcomes one of the primary limitations traditionally associated with analog computing~\cite{chataut_massive_2020,albreem_massive_2019}.

%% file: sections/sec7.tex
\begin{figure}
\centering
\includegraphics[width=\textwidth]{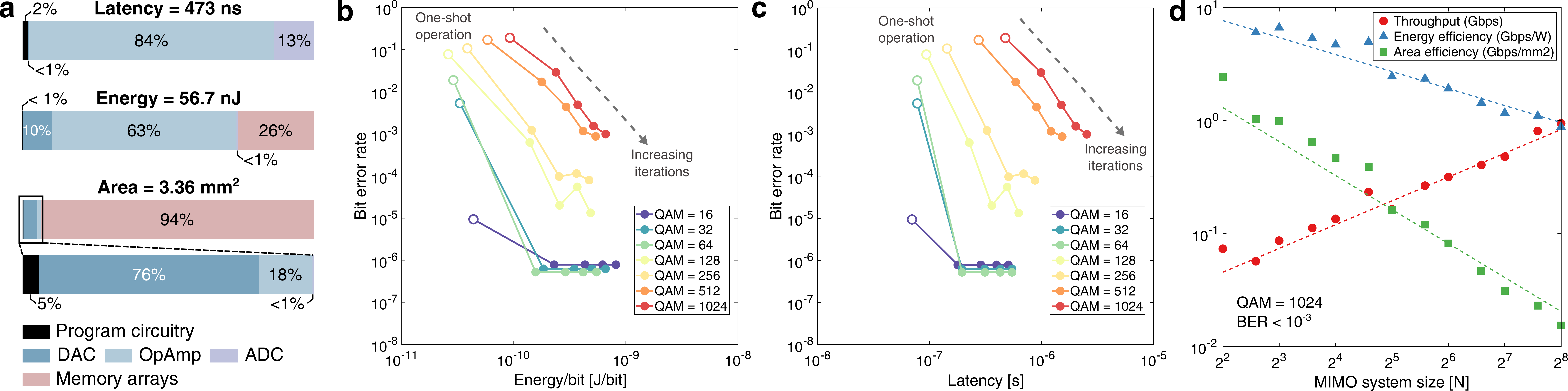}
\caption{\textbf{Benchmarking and scaling.} (a)~Breakdown of energy consumption, latency, and area among memory arrays and peripheral circuitry for a typical single run of the one-shot BCZF solver on a \hl{1024}-QAM 64×64 MIMO decoding problem. (b,c)~BER as a function of (b)~energy-per-bit and (c)~latency for different MIMO modulation orders. Each point corresponds to a different number of iterative refinement steps, illustrating the progressive transition from ultra-low-energy approximate decoding to high-accuracy operation. (d)~Scaling of throughput, energy efficiency, and area efficiency with MIMO size for \hl{1024-QAM} operation at BER $<10^{-3}$. The proposed architecture preserves favorable scaling by combining continuous-time nonlinear optimization with high-precision residual correction.}
\label{fig:bench}
\end{figure}To evaluate the performance of our proposed hybrid analog/digital architecture, we performed benchmark simulations assuming a 14~nm CMOS implementation of both the analog IMC circuit and digital coprocessor (see Methods,~\nameref{meth:benchmark} for additional details). 

Fig.~\ref{fig:bench}a reports the energy, latency, and area breakdown for a representative $64\times64$ \hl{1024}-QAM decoding task. Owing to SRAM-based memory cells that enable fast ($<$\SI{1}{\nano\second}) writing, the programming overhead can be amortized over the channel \hl{reuse interval, assumed here to span} 14 OFDM symbols in 5G-NR systems~\cite{bi_high-speed_2025}. The dominant latency and energy contributions arise from the analog OpAmps, particularly the high-GBWP amplifiers used for local constraint enforcement. Area is instead dominated by the 40T5R memory arrays, which were selected to achieve low analog variability (2\%)~\cite{mannocci_fully_2026}.

Figs.~\ref{fig:bench}b-c report the BER as a function of energy consumption and latency as iterative refinement is progressively applied for modulation orders ranging from 16-QAM to \hl{1024-QAM}. For all constellations, iterative refinement progressively moves the architecture from an ultra-efficient one-shot operating point toward high-accuracy decoding, reducing BER with only moderate increases in energy and latency. Lower-order constellations require fewer refinement iterations, whereas denser constellations continue to benefit from additional correction cycles. Ultimately, the achievable BER is limited by the accuracy of HP residual computation.

Finally, Fig.~\ref{fig:bench}d investigates scalability with increasing MIMO dimension. Throughput, energy efficiency, and area efficiency are reported for operating points achieving BER below $10^{-3}$. Because the number of refinement iterations remains approximately constant with system size, the architecture largely preserves the favorable sub-linear convergence time scaling of the underlying solver, resulting in an equivalent throughput scaling. Energy and area efficiency decrease with size owing to the growth of peripheral circuitry and memory arrays, yet the degradation remains substantially slower than in conventional digital architectures dominated by matrix inversion and iterative optimization. Larger systems can therefore be accommodated through spatial parallelism, with increasing energy per decoded bit as reflected in Fig.~\ref{fig:bench}d.

%% file: sections/sec8.tex
Tab.~\ref{tab:lp_comparison} and Tab.~\ref{tab:hp_comparison} place the proposed architecture within the broader landscape of MIMO detection accelerators by comparing performance on a normalized 64×64 256-QAM task (see Methods,~\nameref{meth:bench_norm}). In the low-precision regime, the one-shot BCZF solver achieves the highest energy efficiency among the considered architectures while maintaining competitive throughput. In the high-precision regime (BER $<$ 10\textsuperscript{-3}), iterative refinement enables competitive throughput and energy efficiency relative to digital and hybrid alternatives. Area efficiency remains limited by the large footprint of the 40T5R memory cells~\cite{mannocci_fully_2026}, although more compact memory technologies such as resistive random access memories (RRAM), phase change memories (PCM), or magnetic random access memories (MRAM)~\cite{sebastian_memory_2020,mannocci_-memory_2023} could substantially improve this metric by a factor 2-6×.

More broadly, the present work suggests a shift in the role of closed-loop in-memory computing. Previous IMC systems have primarily exploited feedback networks to implement equilibrium-based linear algebra operations, including matrix inversion, inverse matrix-vector multiplication, and linear system solving~\cite{sun_solving_2019,sun_one-step_2020,mannocci_generalized_2023,mannocci_fully_2026,zuo_precise_2025,mannocci_iscas_2025}. Here, we instead show that nonlinear optimization can be encoded directly within the continuous-time dynamics of a physical system. In this framework, computation is no longer associated solely with the equilibrium state of the circuit but with the trajectory it follows during its evolution. The proposed BCZF decoder, therefore, serves as a concrete example of a broader paradigm in which continuous-time dynamics become the computational primitive.
\subfile{tables/tab1}%
\subfile{tables/tab2}%

Several limitations nevertheless remain. First, the proposed BCZF architecture has not yet been fabricated as a dedicated integrated circuit. While the underlying computational primitives have been experimentally validated using fabricated IMC hardware and extensive circuit simulations, a full hardware implementation will be required to validate the projected system-level metrics. Second, the present study focuses on a specific nonlinear optimization problem, leaving open the question of how the same principles extend to broader classes of dynamical systems and constrained optimization tasks. Finally, the performance estimates rely on projected models for peripheral circuitry and high-precision residual computation, which will require further validation through future silicon implementations.

These limitations also suggest several directions for future research. Fully integrated implementations combining the nonlinear solver and refinement engine on a single chip could further reduce communication overheads and improve energy efficiency. In addition, scaling beyond the dimensions typically accessible to individual IMC tiles remains an important challenge. Existing IMC systems are constrained by spatial parasitics, interconnect loading, and power-delivery limitations, restricting the practical size of a single computational array. Hierarchical architectures combining multiple nonlinear IMC tiles through high-precision residual correction may provide a route toward substantially larger optimization problems while preserving the efficiency advantages of physical computation. Further opportunities include adaptive refinement strategies that dynamically adjust the number of correction iterations according to the required accuracy, as well as extensions of the proposed framework to a wider range of constrained and unconstrained optimization problems.

By directly encoding a nonlinear problem within the continuous-time dynamics of a feedback network and combining it with mixed-precision iterative refinement, we demonstrate a computational framework that unifies the efficiency of physical optimization with the accuracy of digital correction. Together, these results establish continuous-time nonlinear dynamics as a new computational primitive for in-memory computing and suggest a broader pathway toward scalable, energy-efficient physical optimization systems.

%% file: sections/tables/tab1.tex
\begin{table*}[t]
\centering
\caption{\textbf{Benchmark comparison with existing low-precision MIMO solvers for a 256-QAM 64×64 system.}}
\begin{tabular}{|c|c|c|c|c|c|}
\hline
\rowcolor[cmyk]{0.4,0.1,0,0.4}
\textcolor{white}{\textbf{Metric}} &
\textcolor{white}{\textbf{\cite{tang_058-mm2_2021}}} &
\textcolor{white}{\textbf{\cite{mannocci_analogue_2022}}} &
\textcolor{white}{\textbf{\cite{zuo_precise_2025}}} &
\textcolor{white}{\textbf{\cite{bi_high-speed_2025}}} &
\textcolor{white}{\textbf{This work}} \\
\hline

Architecture &
ASIC &
CL-IMC &
CL+OL-IMC &
CL+OL-IMC &
\textbf{CL-IMC} \\
\hline

Algorithm &
MPD &
MMSE &
ZF &
BCZF (ADMM) &
\textbf{BCZF} \\
\hline

Type &
Iterative &
One-shot &
ZF + IR &
Iterative &
\textbf{One-shot} \\
\hline

BER &
10\textsuperscript{-1} &
10\textsuperscript{-1} &
10\textsuperscript{-1} &
10\textsuperscript{-1} &
\textbf{10\textsuperscript{-1}} \\
\hline

Throughput (Gbps) &
1.38 &
1.23 &
1.55 &
0.205 &
\textbf{\underline{3.6}} \\
\hline

Energy Efficiency (Gbps/W) &
1.563 &
12.3 &
4.85 &
0.074 &
\textbf{\underline{26}} \\
\hline

Energy per bit (pJ/bit) &
639 &
81 &
206 &
13$\cdot$10\textsuperscript{3} &
\textbf{\underline{38}} \\
\hline

Area Efficiency (Mbps/mm\textsuperscript{2}) &
594 &
\underline{2051} &
3.88 &
1.74 &
\textbf{920} \\
\hline

\end{tabular}
\label{tab:lp_comparison}
\end{table*}

%% file: sections/tables/tab2.tex
\begin{table*}[t]
\centering
\caption{\textbf{Benchmark comparison with existing high-precision MIMO solvers for a 256-QAM 64×64 system.}}
\begin{tabular}{|c|c|c|c|c|c|c|}
\hline
\rowcolor[cmyk]{0.4,0.1,0,0.4}
\textcolor{white}{\textbf{Metric}} &
\textcolor{white}{\textbf{\cite{bi_high-speed_2025}}} &
\textcolor{white}{\textbf{\cite{pathak_nvidia_2025}}} &
\textcolor{white}{\textbf{\cite{shahabuddin_admm-based_2021}}} &
\textcolor{white}{\textbf{\cite{tang_058-mm2_2021}}} &
\textcolor{white}{\textbf{\cite{jeon_354_2019}}} &
\textcolor{white}{\textbf{This work}} \\
\hline

Architecture &
FPGA &
GPU &
ASIC &
ASIC &
ASIC &
\textbf{Hybrid} \\
\hline

Algorithm &
ADMM &
ADMM &
ADMM &
MPD &
LAMA &
\textbf{BCZF} \\
\hline

Type &
Iterative &
Iterative &
Iterative &
Iterative &
Iterative &
\textbf{BCZF + IR} \\
\hline

BER &
$<$10\textsuperscript{-3} &
$<$10\textsuperscript{-3} &
$<$10\textsuperscript{-3} &
$<$10\textsuperscript{-3} &
$<$10\textsuperscript{-3} &
\textbf{$<$10\textsuperscript{-3}} \\
\hline

Throughput (Gbps) &
0.144 &
0.228 &
0.377 &
0.97 &
0.354 &
\textbf{\underline{1}} \\
\hline

Energy Efficiency (Gbps/W) &
0.001 &
0.004 &
1.95 &
1.1 &
0.586 &
\textbf{\underline{3.93}} \\
\hline

Energy per bit (pJ/bit) &
776$\cdot$10\textsuperscript{3} &
220$\cdot$10\textsuperscript{3} &
512 &
909 &
1706 &
\textbf{\underline{254}} \\
\hline

Area Efficiency (Mbps/mm\textsuperscript{2}) &
0.058 &
N. A. &
310 &
\underline{418} &
239 &
\textbf{260} \\
\hline

\end{tabular}
\label{tab:hp_comparison}
\end{table*}

%% file: sections/methods.tex
\subsection{Complex to real mapping}
\label{meth:ctor}
The MIMO detection problem under consideration is naturally formulated in the complex domain. However, both the proposed circuit implementation and the experimental IMC platform operate on real-valued quantities. Throughout this work, complex-valued matrices and vectors are therefore mapped to an equivalent real-valued representation.
Given a complex-valued matrix $\Re{\mat{H}}+j\Im{\mat{H}}\in \mathbb{C}^{N_r\times N_t}$, and a complex-valued vector $\Re{\vec{x}}+j\Im{\vec{x}}\in\mathbb{C}^{N_t}$, we define the equivalent real-valued quantities:
\begin{equation*}
\mat{H}_R = 
\begin{bmatrix}
\Re{\mat{H}} & -\Im{\mat{H}}\\
\Im{\mat{H}} & \Re{\mat{H}}
\end{bmatrix}\in \mathbb{R}^{2N_r\times2N_t},
\end{equation*}
\begin{equation*}
\vec{x}_R = \begin{bmatrix}
\Re{\vec{x}} \\
\Im{\vec{x}}
\end{bmatrix}\in \mathbb{R}^{2N_t}.
\end{equation*}
Under this transformation, the Hermitian transpose and matrix-vector multiplication operations map to the equivalent real-valued operations, namely: 
\begin{equation*}
    \mat{H}_R^T = \begin{bmatrix} \Re{\mat{H}}^T & \Im{\mat{H}}^T \\ -\Im{\mat{H}}^T & \Re{\mat{H}}^T \end{bmatrix} = (\mat{H}^\dagger)_R
\end{equation*}
\begin{equation*}
\mat{H}_R\vec{x}_R = \begin{bmatrix} \Re{\mat{H}}\Re{\vec{x}} - \Im{\mat{H}}\Im{\vec{x}} \\\Im{\mat{H}}\Re{\vec{x}} + \Re{\mat{H}}\Im{\vec{x}}\end{bmatrix} = \vec{y}_R,
\end{equation*}
The same representation extends naturally to inverse matrix-vector multiplication. Specifically,
\begin{equation*}
\vec{x} = \mat{H}^{-1}\vec{y} \leftrightarrow \vec{x}_R = \mat{H}_R^{-1}\vec{y}_R
\end{equation*}
As a consequence, all matrix operations appearing in the BCZF formulation—including matrix-vector multiplication, inverse matrix-vector multiplication, Gramian construction, and transpose multiplications—can be implemented using the same real-valued MVM and IMVM primitives. Throughout the paper, reported system dimensions refer to the original complex-valued problem, whereas the underlying hardware operates on the corresponding real-valued representation of dimension $2N_r\times2N_t$.

\subsection{MIMO simulation methodology}
\label{meth:MIMO_simulation}
Unless otherwise specified, channel matrices were generated according to an i.i.d. Rayleigh fading model
\begin{equation*}
h_{ij} \sim \mathcal{CN}(0, \frac{1}{N_r}),
\end{equation*}
such that the average channel coefficient power scales as $1/N_r$~\cite{thrampoulidis_symbol_2018,bjornson_massive_2017}. This normalization maintains approximately constant received signal power as the number of receive antennas varies, enabling a fairer comparison across system sizes, as opposed to the $1$-normalization~\cite{albreem_massive_2019,zheng_massive_2014}. For each channel realization, transmitted vectors were generated by randomly sampling symbols from the target QAM constellation. Let $E_s$ denote the average energy of the unnormalized constellation, \textit{i.e.} $E_s = \mathbb{E}[|s|^2]$. The transmitted symbols were scaled according to
\begin{equation*}
x = \sqrt{\frac{P_{\mathrm{TX}}}{E_sN_t}} s,
\end{equation*}
where $s$ contains the raw QAM symbols and $P_{\mathrm{TX}}$ denotes the total transmit power. This normalization yields:
\begin{equation*}
E[|\mathbf{x}|^2] = P_{\mathrm{TX}},
\end{equation*}
independently of the number of transmit antennas. Unless otherwise specified, $P_{\mathrm{TX}} = 1$.

The received signal was generated as:
\begin{equation*}
\vec{y} = \vec{H}\vec{x} + \vec{w},
\end{equation*}
where:
\begin{equation*}
w_i \sim \mathcal{CN}(0,N_0)
\end{equation*}
represents additive white Gaussian noise. For each channel realization, $N_{\mathrm{tr}} > 10^3$ independent transmitted vectors were generated and decoded. After processing all transmitted vectors, a new independent channel realization was generated, and the procedure was repeated for a total of $N_{\mathrm{ch}} > 20$ channel realizations. Bit-error rates were computed by comparing the detected symbols against the transmitted symbols over the entire ensemble of transmitted vectors and channel realizations.

\subsection{Derivation of the equivalent IMC-BCZF energy function}
\label{meth:energy}
To connect Eq.~\eqref{eq:bczf_circuit_model} to an optimization problem, we define the scalar set function: 
\begin{equation*}
\phi(x) = \frac{k\beta}{2\alpha_0}x^2+\eta_{[-V_s,V_s]}(x),
\end{equation*}
where $\eta_{[-V_s,V_s]}(x)$ is the indicator function of the interval $[-V_s,V_s]$, equal to zero for $x\in[-V_s,V_s]$ and $+\infty$ otherwise. We then consider the energy function:
\begin{equation*}
E(\vec{x}) = \frac{1}{2} \tnormsq{\mat{H}\vec{x}-\vec{y}}+\sum_i \phi(x_i) = \frac{1}{2}\tnormsq{\mat{H}\vec{x}-\vec{y}}+\frac{k\beta}{2\alpha_0}\tnormsq{\vec{x}}+\eta_{[-V_s,V_s]^N}(\vec{x}),
\end{equation*}
which matches the formulation of Eq.~\eqref{eq:bczf_optimality}. A minimizer of the latter satisfies the optimality condition:
\begin{equation*}
0\in\mat{H}^{T}(\mat{H}\vec{x}-\vec{y}) +\partial\sum_i\phi(x_i),
\end{equation*}
where
\begin{equation*}
\partial\phi(x) = \frac{k\beta}{\alpha_0}x+\partial \eta_{[-V_s,V_s]}(x).
\end{equation*}
is the subgradient of $\phi(x)$. 

We now show that this condition is equivalent to the circuit \hl{steady-state} described by Eq.~\eqref{eq:bczf_circuit_model}. For OpAmps operating in the unsaturated regime, $|x_i|<V_s$, the indicator contribution vanishes and the optimality condition yields: 
\begin{equation*}
x_i = -\frac{\alpha_0}{k\beta}[\mat{H}^{T}(\mat{H}\vec{x}-\vec{y})]_i  = \alpha_0 v_i
\end{equation*}
where $v_i$ is the input voltage at the $i$-th lower OpAmp, and which corresponds to the linear region of the amplifier transfer function. For OpAmps at the positive saturation boundary, $x_i=+V_s$, the subgradient of the indicator imposes:
\begin{equation*}
-\frac{\alpha_0}{k\beta}[\mat{H}^{T}(\mat{H}\vec{x}-\vec{y})]_i =\alpha_0 v_i \geq V_s,
\end{equation*}
which is precisely the condition for the amplifier output to saturate at $+V_s$. Similarly, for $x_i=-V_s$, the optimality condition requires saturation at the negative boundary.

Thus, the steady-state circuit solution is equivalent to the minimizer of Eq.~\eqref{eq:bczf_optimality}. The three terms of the energy function correspond respectively to the zero-forcing objective, a weak $\ell_2$ regularization induced by finite amplifier gain, and the box constraint imposed by saturation. In the large-gain limit $\alpha_0\rightarrow\infty$, the regularization term vanishes, and the circuit implements the box-constrained zero-forcing objective.

\subsection{\hl{Continuous-time model of BCZF circuit}}
\label{meth:ct_model}
\hl{To model the BCZF circuit, we start from the Kirchhoff equations at the OpAmp input nodes~\cite{mannocci_generalized_2023}: 
\begin{equation}
\begin{cases}
    \mat{H}_R\vec{v}_{\vec{x}} - \vec{y}_R + k\vec{v}_1 = (k\mat{I}+\mat{U}_{|\mat{H}_R}|)\tilde{\vec{v}}_1 \\
    \mat{H}_R^T\vec{v}_1 = (\mat{V}_{|\mat{H}_R|} + \mat{V}_{\vec{c}})\tilde{\vec{v}}_\vec{x},
\end{cases}
\label{eq:kirch_ct}
\end{equation}
where $\tilde{\vec{v}}_1, \tilde{\vec{v}}_\vec{x}$ are the input voltages at the $\vec{v}_1$ and $\vec{v}_\vec{x}$ OpAmps, respectively, $\mat{U}_{|\mat{H}_R|}$ is the diagonal matrix of row-sums of $|\mat{H}_R|$, $\mat{V}_{|\mat{H}_R|}$ is the diagonal matrix of column-sums of $|\mat{H}_R|$, and $\mat{V}_\mat{c}$ is the diagonal matrix of column sums of $\vec{c}^T$, \textit{i.e.} $\mat{V}_\vec{c} = \diagop{(\vec{c})}$. The choice of $\vec{c}$ outlined in Sec.~\ref{sec:imc_bczf}, allows to reduce the diagonal scaling on $\tilde{\vec{v}}_\vec{x}$ to a scalar $\beta$ as:
\begin{equation*}
    \mat{V}_{|\mat{H}_R|} + \mat{V}_\vec{c} = \max\big(\sum_i|H_{R,ij}|\big)\mat{I} = \beta\mat{I}.
\end{equation*}
Kirchhoff equations are coupled with the continuous-time equations for the OpAmps, which are modeled using a single-pole approximation as~\cite{mannocci_generalized_2023}:
\begin{equation*}
    \begin{cases}
        -G_m\tilde{\vec{v}}_1 &= g_0\vec{v}_1 + C_0\frac{d\vec{v}_1}{dt}\\ 
        G_m\tilde{\vec{v}}_\vec{x} &= g_0\vec{v}_\vec{x} + C_0\frac{d\vec{v}_\vec{x}}{dt} + i(\vec{v}_\vec{x}).  \\ 
    \end{cases}
\end{equation*}
Introducing the open-loop DC gain $\alpha_0 = G_m/g_0$, the gain-bandwidth product $p_0 = G_m/C_0$ and $f(\vec{v}) = i(\vec{v})/G_m$, the latter can be rewritten as: 
\begin{equation}
    \begin{cases}
        -\tilde{\vec{v}}_1 &= \frac{1}{p_0}\frac{d\vec{v}_1}{dt} +\frac{1}{\alpha_0}\vec{v}_1 \\ 
        \tilde{\vec{v}}_\vec{x} &= \frac{1}{p_0}\frac{d\vec{v}_\vec{x}}{dt} +\frac{1}{\alpha_0}\vec{v}_\vec{x} + f(\vec{v}_\vec{x}).  \\ 
    \end{cases}
    \label{eq:opamps_ct}
\end{equation}
Coupling Eqs.~(\ref{eq:kirch_ct}-\ref{eq:opamps_ct}) thus yields the dynamical system: 
\begin{equation*}
    \begin{cases}
        \mat{H}_R\vec{v}_{\vec{x}} - \vec{y}_R + k\vec{v}_1 = -(k\mat{I}+\mat{U}_{|\mat{H}_R|})\big(\frac{1}{p_0}\frac{d\vec{v}_1}{dt} +\frac{1}{\alpha_0}\vec{v}_1\big) \\
    \mat{H}_R^T\vec{v}_1 = \beta\big(\frac{1}{p_0}\frac{d\vec{v}_\vec{x}}{dt} +\frac{1}{\alpha_0}\vec{v}_\vec{x} + f(\vec{v}_\vec{x})\big),
    \end{cases}
\end{equation*}
which can be finally rewritten as:
\begin{equation*}
    \begin{cases}
        -\mat{U}_1^{-1}\big[(k\mat{I}+\frac{\mat{U}_1}{\alpha_0})\vec{v}_1 + \mat{H}_R\vec{v}_{\vec{x}} - \vec{y}\big] = \frac{1}{p_0}\frac{d\vec{v}_1}{dt} \\
    \frac1\beta \mat{H}_R^T\vec{v}_1 - \frac{1}{\alpha_0}\vec{v}_\vec{x} - f(\vec{v}_\vec{x}) = \frac{1}{p_0}\frac{d\vec{v}_\vec{x}}{dt},
    \end{cases}
\end{equation*}
where $\mat{U}_1 = (k\mat{I}+\mat{U}_{|\mat{H}_R|})$. Assuming $k\mat{I} \gg \frac{\mat{U}_1}{\alpha_0}$ finally yields the continuous-time state equations for the BCZF circuit: 
\begin{equation}
    \begin{cases}
      \frac{d\vec{v}_1}{dt} \simeq -p_0\mat{U}_1^{-1}\big(k\vec{v}_1 + \mat{H}_R\vec{v}_{\vec{x}} - \vec{y}_R\big)\\
      \frac{d\vec{v}_\vec{x}}{dt} \simeq p_0\big(\frac1\beta\mat{H}_R^T\vec{v}_1 - \frac{1}{\alpha_0}\vec{v}_\vec{x} - f(\vec{v}_\vec{x})\big) ,
    \end{cases}
    \label{eq:bczf_ct_model}
\end{equation}
corresponding to Eqs.~(\ref{eq:ode_1}-\ref{eq:ode_2}).}

\subsection{Convergence time analysis of nonlinear BCZF circuit}
\label{meth:tts}
\hl{For a stable, linear closed-loop IMC circuit programmed with a matrix $\mat{M}$, the convergence time, \textit{i.e.} the time required for output voltages to reach their steady-state condition, is determined by the slowest pole, given by: 
\begin{equation}
    p_{min} = p_0 \lambda_{min} = 1/\tau
\end{equation}
where $\tau$ is the resulting closed-loop circuit time constant, $p_0$ is the OpAmp gain-bandwidth product, and $\lambda_{min}$ is the smallest eigenvalue of the voltage partition matrix $\mat{U}^{-1}\mat{M}$, where $\mat{U}$ is the diagonal matrix of row-sums of $\mat{M}$, plus any loading matrix external to the feedback matrix~\cite{sun_time_2020,mannocci_generalized_2023}. For complex-valued eigenvalues, the quantity of interest is the minimal real part across $\lambda$. For the nonlinear BCZF circuit, the equivalent $\mat{M}$ and $\mat{U}$ matrices are given by: 
\begin{equation*}
    \mat{M} = \begin{bmatrix}k\mat{I} & \mat{H}_R \\ -\mat{H}_R^T & \mat{0} \end{bmatrix}, \ \ \ \ \mat{U} = \begin{bmatrix} \mat{U}_1 & \mat{0} \\ \mat{0} & \beta\mat{I} \end{bmatrix}.
\end{equation*}
If the circuit were to operate linearly, \textit{i.e.} in absence of supply voltage limitations, the size-scaling of the \textit{analog} convergence time, \textit{i.e.} the convergence time of analog voltages $\vec{v}_\vec{x}$, would approximately be determined by the size-scaling of the minimal eigenvalue of $\mat{U}^{-1}\mat{M}$, yielding an $\mathcal{O}(N^2)$ dependence for Gaussian channel matrices (Extended Data Fig.~\ref{exfig:tts}a,b), and no dependence on the QAM depth $M$. However, it suffices for voltages $\vec{v}_\vec{x}$ to fall within the decision boundaries of the constellation to produce constant quantized outputs $\tilde{\vec{x}}_{\textrm{IMC}}$, irrespective of residual analog dynamics. As such, post-ADC convergence time typically shows an additional dependence on $M$ (Extended Data Fig.~\ref{exfig:tts}c). Given the exponential nature of the analog transient~\cite{mannocci_generalized_2023}, and since decision boundaries progressively shrink as $1/\sqrt{M}$~\cite{spagnolini_statistical_2018}, the resulting dependence is of order $\mathcal{O}(\log(M))$, such that the post-ADC convergence time for a fully-linear circuit shows a composite dependence $\mathcal{O}(N^2 \log(\sqrt{M}))$.}

\hl{However, the saturation nonlinearity introduces a pole-movement effect which dynamically alters the circuit time-constant during the transient. Extended Data Fig.~\ref{exfig:tts}d schematically shows the equivalent, locally-linear circuit obtained from a generic closed-loop circuit with input matrix $\mat{Y}$, feedback matrix $\mat{X}$, input currents $\vec{i}_{in}$ and input voltages $\vec{v}_{in}$, when a subset of the OpAmps saturates, causing their feedback loops to be interrupted. Saturated OpAmps can be equivalently replaced with voltage sources $\vec{v}_{sat}$, providing a constant-voltage output at either $+V_{s}$ or $-V_{s}$, and with an open circuit at the corresponding crosspoint rows. In the equivalent circuit, only a submatrix $\mat{X}_{LL}$ of the original feedback matrix $\mat{X}$ still participates in the feedback action, thus resulting in a new dominant time constant $\tau$. Conversely, submatrix $\mat{X}_{LS}$ behaves as an equivalent input array driven by voltages $\vec{v}_{sat}$, together with the submatrix $\mat{Y}_L$ of the input array. Finally, submatrices $\mat{Y}_S$, $\mat{X}_SS$ and $\mat{X}_SL$ do not participate in the active linear feedback subsystem, but determine the validity of the local linearization by influencing the equivalent input voltage $\vec{v}_{chk}$ of the saturated OpAmps.}      

\hl{Extended Data Fig.~\ref{exfig:tts}e shows an example transient for a 64×64, 16-QAM circuit, highlighting a progressive reduction of the closed-loop time constant and corresponding acceleration of the circuit dynamics with respect to a fully-linear equivalent circuit. As such, we model the convergence time $T_{NL}(N,M)$ for the nonlinear BCZF circuit through a saturation-induced speedup factor $S(N,M)$: 
\begin{equation}
    T_{NL}(N,M) = \frac{T_{L}(N,M)}{S(N,M)} = \frac{\tau_L N^2 \log(\sqrt{M})}{1+\gamma P_{edge}(N,M)q(N,M)^\eta},
    \label{eq:tts_fit}
\end{equation}
where $T_{L}(N,M)$ is the convergence time for the equivalent fully-linear circuit, and $\tau_L,\gamma,\eta$ are fitting parameters representing the linear-circuit time-constant, saturation speedup coefficient, and saturation speedup scaling with the number of boundary components, respectively. $P_{edge}(N,M)$ is the probability that the decoded vector contains at least one boundary symbol, and $q(N,M)$ is the expected total number of components lying on the constellation boundary, considering both in-quadrature and in-phase components. $P_{edge}(N,M)$ and $q(N,M)$ thus act as statistical proxies for saturation activity in the fitted model. For square-QAM constellations, it holds: 
\begin{equation*}
    P_{edge}(N,M) = 1 - \bigg(1-\frac{2}{\sqrt{M}}\bigg)^{2N}\ \ \ \ \ q(N,M) = 4\frac{N}{\sqrt{M}}.
\end{equation*}
The corresponding asymptotic convergence time at sufficiently large $N$ is thus: 
\begin{equation*}
T_{NL}(N,M) \sim \frac{\tau_L N^2 \log(\sqrt{M})}{\gamma \ 4^\eta \frac{N^\eta}{M^{\frac{\eta}{2}}}} \simeq \frac{\tau_L}{2\gamma\ 4^\eta}N^{2-\eta}\ M^{\frac{\eta}{2}}\ \log(M).
\end{equation*}
Least-squares fitting particularly yields $\tau_L = \SI{3.7}{\nano\second}$, $\gamma = 0.04$ and $\eta \simeq 2$ (Extended Data Fig.~\ref{exfig:tts}f), thus resulting in approximately size-independent behavior for sufficiently large $N$.}

\subsection{Linear algebra IMC testchip and experimental platform}
\label{meth:platform}
The IC was designed in ST Microelectronics 90~nm CMOS technology around two memory arrays of 4096 devices each. Memory arrays were organized as matrices of 64 word lines (WLs) and 64 bit lines (BLs), with a 40T5R cell at each WL-BL intersection, where resistors were realized using lightly-doped n-well structures. WL/BL pairs can be connected to dedicated direct memory access (DMA) pins, allowing array characterization by an external circuit. In addition to memory devices, the chip integrates OpAmps, DACs, and ADCs. Opamps were designed using a two-stage folded cascode topology with Miller compensation, achieving an open-loop gain of 66~dB and 20~MHz bandwidth. DACs were designed in a 7-bit R-2R topology, where an additional OpAmp buffered the generated analog value. The comparator-type ADC exploits an 8-bit shared staircase, which is fed as a reference to one comparator per output voltage. Each comparator is connected to an independent register, which latches the content of the staircase register when the corresponding output voltage falls below the staircase voltage. Communication to and from the chip is performed using an 8-bit parallel I/O bus, allowing access to on-chip configuration registers.     

The IC was interfaced to a hardware platform comprising a custom printed circuit board (PCB) hosting an STM32F429ZI microcontroller unit (MCU) with custom firmware. The PCB features a serial-to-USB converter to allow interfacing with an external computer running MATLAB. The firmware comprised an internal finite-state machine (FSM) allowing execution of pre-programmed routines for matrix programming, circuit reconfiguration, data transmission to the IC, and fetching of IMC results from the IC. Additionally, the PCB comprised an on-board characterization circuit (OBCC), composed of a 10-bit test voltage generator, a high-side current sense circuit with programmable gain, and a 12-bit ADC, which can be connected to the IC DMA pins through an on-board switching matrix. All power supplies required for IC, MCU, and OBCC operation, namely 1.8~V, 0.9~V, 3.3~V, and 5~V, were generated from an external 7.5~V supply using low-dropout regulators. 

\subsection{Continuous-time emulation methodology}
\label{meth:ct_emulation}
The nonlinear decoder introduced in this work is fundamentally described by a continuous-time dynamical system. To experimentally validate its behavior using the available IMC test chip, the continuous-time dynamics were first transformed into an equivalent discrete-time formulation that could be executed through repeated MVM and IMVM operations.

Starting from the continuous-time model of Eqs.~(\ref{eq:ode_1}-\ref{eq:ode_2}), and adding an extra $\frac{1}{\alpha_0}$ term in the first block to account for the finite gain of upper OpAmps, we obtain:
\begin{equation*}
    \frac{1}{p_0}\frac{d}{dt}\begin{bmatrix}\vec{v}_1\\\vec{v}_\vec{x}\end{bmatrix} 
    \simeq 
    \begin{bmatrix}
    -\hat{\mat{K}} -\frac{1}{\alpha_0}\mat{I} & -\hat{\mat{H}}_R \\ \frac{1}{\beta}\mat{H}_R^T & -\frac{1}{\alpha_0}\mat{I}
    \end{bmatrix}
    \begin{bmatrix}\vec{v}_1\\\vec{v}_\vec{x}\end{bmatrix}
    -\begin{bmatrix}\vec{0}\\f(\vec{v}_\vec{x})\end{bmatrix}
    + \begin{bmatrix}\mat{U}_1^{-1}\vec{y}_R\\ \vec{0}\end{bmatrix}
\end{equation*}
where $\hat{\mat{K}} = \mat{U}_1^{-1}k$ and $\hat{\mat{H}}_R = \mat{U}_1^{-1}\mat{H}_R$ are the voltage divider matrices over conductances $k$ and matrix $\mat{H}_R$, respectively~\cite{mannocci_generalized_2023}. First-order discretization of the derivative terms yields the equivalent discrete-time model: 
\begin{equation*}
       \frac{1}{p_0}\frac{1}{\Delta t}\begin{bmatrix}\vec{v}_1[t] - \vec{v}_1[t-\Delta t]\\\vec{v}_\vec{x}[t] - \vec{v}_\vec{x}[t-\Delta t]\end{bmatrix} 
    \simeq 
    \begin{bmatrix}
    -\hat{\mat{K}} -\frac{1}{\alpha_0}\mat{I} & -\hat{\mat{H}}_R \\ \frac{1}{\beta}\mat{H}_R^T & -\frac{1}{\alpha_0}\mat{I}
    \end{bmatrix}
    \begin{bmatrix}\vec{v}_1[t]\\\vec{v}_\vec{x}[t]\end{bmatrix}
    -\begin{bmatrix}\vec{0}\\f(\vec{v}_\vec{x}[t])\end{bmatrix}
    + \begin{bmatrix}\mat{U}_1^{-1}\vec{y}_R[t]\\ \vec{0}\end{bmatrix},
\end{equation*}
which can be rearranged as: 
\begin{equation*}
    \Bigg(\bigg(\frac{1}{p_0 \Delta t} + \frac{1}{\alpha_0}\bigg)\mat{I}      +\begin{bmatrix}
    \hat{\mat{K}} & \hat{\mat{H}}_R \\ -\frac{1}{\beta}\mat{H}_R^T & \vec{0}
    \end{bmatrix}\Bigg)   
     \begin{bmatrix}\vec{v}_1[t]\\\vec{v}_\vec{x}[t]\end{bmatrix}
     \simeq  
    \frac{1}{p_0 \Delta t} \begin{bmatrix}\vec{v}_1[t-\Delta t]\\\vec{v}_\vec{x}[t-\Delta t]\end{bmatrix}
    -\begin{bmatrix}\vec{0}\\f(\vec{v}_\vec{x}[t])\end{bmatrix}
    + \begin{bmatrix}\mat{U}_1^{-1}\vec{y}_R[t]\\ \vec{0}\end{bmatrix},
\end{equation*}
Multiplying by $p_0\Delta t$ yields: 
\begin{equation*}
    \Bigg(\bigg(1 + \frac{p_0 \Delta t}{\alpha_0}\bigg)\mat{I}      +p_0 \Delta t\begin{bmatrix}
    \hat{\mat{K}} & \hat{\mat{H}}_R \\ -\frac{1}{\beta}\mat{H}_R^T & \vec{0}
    \end{bmatrix}\Bigg)   
    \begin{bmatrix}\vec{v}_1[t]\\\vec{v}_\vec{x}[t]\end{bmatrix}
     \simeq  
    \begin{bmatrix}\vec{v}_1[t-\Delta t]\\\vec{v}_\vec{x}[t-\Delta t]\end{bmatrix}
    -p_0\Delta t\begin{bmatrix}\vec{0}\\f(\vec{v}_\vec{x}[t])\end{bmatrix}
    + p_0\Delta t\begin{bmatrix}\mat{U}_1^{-1}\vec{y}_R[t]\\ \vec{0}\end{bmatrix},
\end{equation*}
which can be compactly rewritten by suitably defining matrices $\mat{A},\mat{B}$ and vectors $\vec{v},\vec{i}$ as:
\begin{align*}
\mat{A}\vec{v}[t] = 
    \mat{A}\begin{bmatrix}\vec{v}_1[t]\\\vec{v}_\vec{x}[t]\end{bmatrix} 
    &\simeq 
    \begin{bmatrix}\vec{v}_1[t-\Delta t]\\\vec{v}_\vec{x}[t-\Delta t]\end{bmatrix}
    -p_0\Delta t\begin{bmatrix}\vec{0}\\f(\vec{v}_\vec{x}[t])\end{bmatrix}
    +\mat{B}\begin{bmatrix}\vec{y}_R[t]\\ \vec{0}\end{bmatrix} \\
    &=
    \vec{v}[t-\Delta t] 
    -p_0\Delta t\begin{bmatrix}\vec{0}\\f(\vec{v}_\vec{x}[t])\end{bmatrix}
    +\mat{B}\vec{i}[t]
    \numberthis\label{eq:dt_model_pre}
\end{align*}

To enable the computation of the DT model using our MVM/IMVM hardware, we use a two-step scheme consisting of (i)~unconstrained state computation and (ii)~saturated projection. We first define the unconstrained state $\vec{v}'[t]$ obtained by neglecting the nonlinear saturation term during one update step:
\begin{equation*}
    \mat{A}\vec{v}'[t] = \vec{v}[t-\Delta t]+\mat{B}\vec{i}[t].
\end{equation*}
Subtracting this equation from the full implicit update gives:
\begin{equation}
    \mat{A}(\vec{v}[t]-\vec{v}'[t]) = -p_0\Delta t \begin{bmatrix}\vec{0}\\f(\vec{v}_{x}[t])\end{bmatrix}.
    \label{eq:ode_sub}
\end{equation}
For sufficiently small $\Delta t$, we can approximate: 
\begin{equation*}
    \frac{1}{p_0\Delta t}\mat{A} \simeq \Bigg(\bigg(\frac{1}{p_0\Delta t} + \frac{1}{\alpha_0}\bigg)\mat{I}      +\begin{bmatrix}
    \hat{\mat{K}} & \hat{\mat{H}}_R \\ -\frac{1}{\beta}\mat{H}_R^T & \vec{0}
    \end{bmatrix}\Bigg) \simeq \frac{1}{p_0\Delta t}\mat{I},
\end{equation*}
such that Eq.~\eqref{eq:ode_sub} can be approximated to the system: 
\begin{equation}
    \begin{cases}
        \vec{v}_1[t] - \vec{v}'_1[t] = \vec{0} \\
        \vec{v}_{\vec{x}}[t]-\vec{v}'_\vec{x}[t] \in -f(\vec{v}_{x}[t]),
    \end{cases}
\end{equation}
where the scalar factor $p_0\Delta t$ has been absorbed in the definition of $f(\cdot)$ for compactness, and $\in$ is used owing to $f(\cdot)$ being a set-value map. Recalling the definition of $f(x)$, namely $f(x) = 0$ for $x\in (-V_s,V_s)$, $f(x) \in [0,+\infty)$ for $x = V_s$ and $f(x) \in (-\infty,0]$ for $x = -V_s$, the inverse map relation yields: 
\begin{equation*}
    \vec{v}_{\vec{x}}[t] \in f^{-1}(\vec{v}'_{x}[t]-\vec{v}_{x}[t]),
\end{equation*}
where $f^{-1}(x) \in [-V_s,V_s]$ for $x = 0$, $f^{-1}(x) = V_s$ for $x > 0$ and $f^{-1}(x) = -V_s$ for $x < 0$. The latter is equivalent to applying an (elementwise) clipping saturation between $-V_s$ and $V_s$ to the unconstrained output $\vec{v}'_\vec{x}[t]$:
\begin{equation}
    \begin{cases}
        \vec{v}_{\vec{x}}[t] = V_s & \leftrightarrow \vec{v}'_{x}[t]-\vec{v}_{x}[t] > 0 \\
        \vec{v}_{\vec{x}}[t] \in [-V_s,V_s] & \leftrightarrow \vec{v}'_{x}[t]-\vec{v}_{x}[t] = 0 \\
        \vec{v}_{\vec{x}}[t] = -V_s & \leftrightarrow \vec{v}'_{x}[t]-\vec{v}_{x}[t] < 0
    \end{cases}
    \Leftrightarrow
    \begin{cases}
        \vec{v}_{\vec{x}}[t] = V_s & \leftrightarrow \vec{v}'_{x}[t] \geq V_s \\
        \vec{v}_{x}[t] = \vec{v}'_{x}[t]& \leftrightarrow |\vec{v}'_{\vec{x}}[t]| \leq V_s \\
        \vec{v}_{\vec{x}}[t] = -V_s & \leftrightarrow \vec{v}'_{x}[t] \leq -V_s
    \end{cases}
    \Leftrightarrow \operatorname{clip}(\vec{v}'_{x}[t],-V_s,V_s), 
\end{equation}
which corresponds to the second step of the two-step update process. Intuitively, the two-step update can be regarded as modeling a round-trip propagation through the feedback loop from $\vec{v}_{\vec{x}}[t-\Delta t]$ to the ideal, unconstrained $\vec{v}'_{\vec{x}}[t]$, followed by application of the saturating action of the supply-limited OpAmps yielding the updated $\vec{v}_{\vec{x}}[t]$:
\begin{equation}
    \begin{cases}
        \vec{v}'[t] &= \mat{A}^{-1}\big(\vec{v}[t-\Delta t]+\mat{B}\vec{i}[t]\big) \\
        \vec{v}_{\vec{x}}[t] &= \operatorname{clip}(\vec{v}'_{x}[t],-V_s,V_s)            
        \label{eq:dt_2step}
    \end{cases}
\end{equation}
As shown in Extended Data Fig.~\ref{exfig:ctvsdt}, Eq.~\eqref{eq:dt_2step} enables accurate reproduction of the continuous-time trajectories for both full channel precision (FP64) and quantized channels (INT5), as long as $\Delta t$ is chosen to be sufficiently small (100~ps-1~ns), while deviations appear at larger update steps (100~ns) because the finite propagation delay no longer approximates the continuous feedback dynamics.

A direct implementation of the equivalent matrix $\mat{A}$ is, however, problematic because its different matrix blocks exhibit substantially different numerical ranges (see Extended Data Fig.~\ref{exfig:dtmat}). Recalling the definition of $\mat{A}$, 
\begin{equation*}
    \mat{A} = p_0\Delta t\begin{bmatrix}
    \big(\frac{1}{p_0 \Delta t}+\frac{1}{\alpha_0}\mat{I}\big) + \hat{\mat{K}} & \hat{\mat{H}}_R \\
    -\frac{1}{\beta}\mat{H}_R^T     & \big(\frac{1}{p_0\Delta t} + \frac{1}{\alpha_0}\big)\mat{I}
    \end{bmatrix},
\end{equation*}
off-diagonal blocks scale as $\Delta t$, whereas diagonal blocks do not scale with $\Delta t$, thus potentially leading to excessive quantization error when the matrix is mapped onto the finite-precision hardware. To mitigate this issue, $\mat{A}$ was partitioned as:
\begin{equation*}
\mat{A} = \begin{bmatrix}
\mat{A}_1 & \mat{A}_3\\
\mat{A}_2 & \mat{A}_4
\end{bmatrix} 
\end{equation*}
and solved using a block-IMVM approach based on Schur-complement decomposition~\cite{pan_blockamc_2024,mannocci_fully_2026}. Specifically, the linear system was reformulated as a sequence of three IMVM operations and two MVM operations:
\begin{align*}
\vec{x}_1' &= \mat{A}_1^{-1}\vec{y}_1,\\
\vec{y}_2' &= \mat{A}_2\vec{x}_1',\\
\vec{x}_2 &= \mat{S}^{-1}(\vec{y}_2-\vec{y}_2'),\\
\vec{y}_1' &= \mat{A}_3\vec{x}_2,\\
\vec{x}_1 &= \mat{A}_1^{-1}(\vec{y}_1-\vec{y}_1'),
\end{align*}
where:
\begin{equation*}
\mat{S} =  \mat{A}_4 - \mat{A}_2\mat{A}_1^{-1}\mat{A}_3
\end{equation*}
denotes the Schur complement.
\hl{Notably, the testchip used in this work can natively execute IMVM operations up to matrices of size 64×64~\cite{mannocci_fully_2026}. Conversely, MVM operations with a target matrix $\mat{B}\in\mathbb{R}^{m\times n}$ require the definition of the augmented linear system: 
\begin{equation*}
    \begin{bmatrix}\mat{I} & \mat{0}\\\mat{B} & \mat{I}\end{bmatrix}^{-1}    \begin{bmatrix}\vec{x}\\\vec{0}\end{bmatrix} = \begin{bmatrix}\vec{x}\\-\mat{B}\vec{x}\end{bmatrix},
\end{equation*}
where the resulting block-matrix has size $(n+m) \times (n+m)$.}
Finally, the nonlinear mapping $f^{-1}(\cdot)$ was implemented digitally after each update step. This procedure effectively reproduces the evolution of the proposed physical dynamical system while naturally incorporating all hardware nonidealities of the experimental platform.

\hl{\subsection{Experimental validation and die-to-die variability}
\label{meth:expdataset}
For an  $N\times N$ complex-valued channel $\mat{H}$, the equivalent real-valued channel matrix $\mat{H}_R$ has size $2N\times 2N$, while the corresponding circuit matrix $\mat{A}$ has dimensions $4N\times4N$. Block-IMVM thus requires execution of IMVM and MVM operations on blocks of size 
$2N\times2N$, with each intermediate MVM implemented through an augmented linear system of size $4N\times4N$. Consequently, the largest channel that can be emulated on the test chip is 16×16: its 32×32 real-valued representation requires a 64×64 augmented linear system, thereby occupying the complete memory array. 
Experiments were performed for random 4×4, 8×8, and 16×16 MIMO channel realizations using a discretization step of $\Delta t = \SI{1}{\nano\second}$ and were repeated on three dies to assess device-to-device variability. At time step i, agreement between the experimental and ideal voltage vectors is quantified by:
\begin{equation}
    \varepsilon[i] = \frac{\tnorm{\vec{v}_{\vec{x},exp}[i] - \vec{v}_{\vec{x},id}[i]}}{\tnorm{\vec{v}_{\vec{x},id}[i]}},
\end{equation}
where $\vec{v}_{exp}[i]$ and $\vec{v}_{id}[i]$ are the experimental and ideal $\vec{v}_\vec{x}$ voltage vectors, respectively. Extended Data Fig.~\ref{exfig:expaccuracy}a shows the distributions aggregated across dies for each MIMO dimension. The different dimensions exhibit comparable error distributions, with median errors of approximately 2\% and standard deviations of approximately 0.7 percentage points. Extended Data Fig.~\ref{exfig:expaccuracy}b shows the distributions aggregated across channel dimensions for each die. The die-specific median errors differ by approximately 0.4 percentage points, indicating limited die-to-die variability. Extended Data Fig.~\ref{exfig:expaccuracy}c summarizes the results for every channel-dimension and die combination, further supporting the accuracy and reproducibility of the experimental emulation procedure.}

\subsection{Hardware assumptions and performance projection}
\label{meth:benchmark}
The performance projections reported in this work were obtained using a bottom-up hardware model based on published 14 nm CMOS building blocks. The proposed decoder comprises three main components: a nonlinear closed-loop IMC solver, peripheral conversion and programming circuitry, and a high-precision matrix-vector multiplication (HP-MVM) unit used for iterative refinement \hl{and channel-dependent processing}.

The nonlinear solver was modeled using 5-bit differential resistive memory arrays storing the channel matrix together with operational amplifiers implementing the feedback dynamics. Memory-cell area was derived from the 40T5R architecture reported in~\cite{mannocci_fully_2026}, scaled to a 14~nm technology node. Programming circuitry was modeled using SRAM-based write drivers, while current-input DACs, voltage-output DACs, and shared-ramp ADCs were derived from previously reported 14~nm implementations~\cite{kim_design_2019,mannocci_fully_2026}. ADC energy, latency, and area were scaled according to resolution following the methodology of~\cite{marinella_multiscale_2018,murmann_mixed-signal_2021}. Operational amplifiers were modeled using the design reported by Feinberg \textit{et al.}~\cite{feinberg_analog_2021}, with gain-bandwidth products selected to satisfy the convergence and accuracy requirements of the nonlinear solver and subsequently scaled power consumption following the methodology described in~\cite{zuo_precise_2025}.

The HP-MVM unit was modeled separately from the analog solver \hl{using the digital matrix-vector multiplication accelerator reported in~\cite{zimmer_011_2019}}. The reported energy cost of 0.105 pJ per operation at minimum clock frequency was scaled according to the operating frequency required to satisfy the target latency. Similarly, throughput density was obtained from the reported throughput-per-area figures and used to derive the required silicon area for the residual engine. \hl{The same HP-MVM hardware was assumed to perform both the residual evaluations required by iterative refinement and the preprocessing required to calculate the channel-dependent circuit parameters $k$ and $\vec{c}$.} 

For a problem of \hl{complex} dimension $N$, the total number of HP-MVM operations was assumed to scale as:
\begin{equation*}
N_{\mathrm{op,MVM}} = 2 \cdot 4N^2 (N_{\mathrm{iter}}-1),
\end{equation*}
where the factor of four accounts for the equivalent real-valued representation of complex matrices and the factor of two converts multiply-accumulate operations into elementary arithmetic operations. \hl{The channel-dependent preprocessing required to calculate $k$ and $\vec{c}$ was assigned an operation count:
\begin{equation*}
N_{\mathrm{op,pre}} = 6N^2+N-2.
\end{equation*}}%
\hl{We assume a 5G-NR subcarrier spacing (SCS) of \SI{120}{\kilo\hertz}, corresponding to a useful OFDM-symbol interval $T_{sym} = 1/SCS = \SI{8.33}{\micro\second}$, and consider the channel matrix to be updated once per 14-symbol slot. Channel preprocessing and programming are therefore performed once per channel update and amortized over $N_{reuse} = 14$ decoded vectors. For an HP-MVM throughput $R_{\mathrm{MVM}}$, the residual-computation and preprocessing latencies thus read: 
\begin{equation*}
    T_{\mathrm{MVM}} = \frac{N_{\mathrm{op,MVM}}}{R_{\mathrm{MVM}}},
\end{equation*}
\begin{equation*}
    T_{\mathrm{pre}} = \frac{N_{\mathrm{op,pre}}}{R_{\mathrm{MVM}}}.
\end{equation*}
The equivalent per-vector decoder latency was consequently evaluated as: 
\begin{equation*}
    T_{\mathrm{dec}} =T_{\mathrm{IMC}} + T_{\mathrm{MVM}} + \frac{T_{\mathrm{prog}} + T_{\mathrm{pre}}}{N_{reuse}},
\end{equation*}
where $T_{\mathrm{IMC}}$ includes the analog computing latency accumulated over all iterative-refinement steps, and $T_{\mathrm{prog}}$ is the latency required to program the channel matrix and channel-dependent circuit quantities.}

\hl{For each MIMO dimension, the HP-MVM accelerator’s tile count was selected to meet the latency requirements of the most demanding supported combination of modulation order and refinement count. The resulting throughput and area were then held fixed across all operating points at that dimension. The accelerator area was calculated from its throughput density as ($A_{\mathrm{MVM}}=R_{\mathrm{MVM}}/\rho_{\mathrm{MVM}}$). The selected throughput satisfies the conservative symbol-time constraint $T_{\mathrm{dec}}\leq T_{\mathrm{sym}}$, while also ensuring a complete-decoder throughput of at least 1~Gbps at the 256-QAM operating point achieving $\mathrm{BER} < 10^{-3}$.}

\hl{The equivalent energy required to decode one vector was calculated as: 
\begin{equation*}
    E_{\mathrm{dec}} = E_{\mathrm{IMC}} + E_{op}\ N_{\mathrm{op,MVM}} + \frac{E_{\mathrm{prog}} + E_{op}\ N_{\mathrm{op,pre}}}{N_{reuse}},
\end{equation*}
where $E_{op} = \SI{1.3}{\pico\joule}$ is the energy cost per-operation~\cite{zimmer_011_2019}, $E_{\mathrm{IMC}}$ includes the analog computing energy accumulated over all iterations, and  $E_{\mathrm{prog}}$ is the channel-programming energy.}

\hl{Each decoded vector contains $N_{bit} = N \log_2(M)$ information bits for $M$-QAM modulation. Throughput, energy per bit, and energy efficiency were therefore calculated as:
\begin{equation*}
    R_{bit} = \frac{N_{bit}}{T_{\mathrm{dec}}}, \ \ \ \ \ E_{bit} = \frac{E_{\mathrm{dec}}}{N_{bit}}, \ \ \ \ \ \eta_{E} = \frac{1}{E_{bit}}.
\end{equation*}
The complete-decoder area was calculated as: 
\begin{equation}
    A_{\mathrm{dec}} = A_{\mathrm{IMC}} + A_{MVM} = A_{\mathrm{IMC}} + \frac{R_{\mathrm{MVM}}}{\rho_{\mathrm{MVM}}},
\end{equation}
where $\rho_{\mathrm{MVM}}$ is the throughput density of the digital residual engine~\cite{zimmer_011_2019}.}

Unless otherwise stated, all reported results assume a 14 nm implementation, a supply voltage of 1.2 V, a 5-bit analog memory representation, 8-bit high-precision residual computation, and five iterative refinement steps. Energy, latency, throughput, energy efficiency, and area efficiency were obtained by combining the contributions of the nonlinear solver, peripheral circuitry, and HP-MVM engine according to the methodology described above.

\subsection{Benchmark normalization}
\label{meth:bench_norm}
Published MIMO detection accelerators target a wide range of channel dimensions, modulation formats, and error-rate requirements, making direct comparison challenging. To facilitate a consistent evaluation, all reported metrics were normalized whenever possible to a common $64\times64$ 256-QAM benchmark. Normalization was performed using the algorithmic complexity reported in the original references together with scaling laws derived from the corresponding architectures.

For fully programmable platforms such as GPUs and FPGAs, latency was estimated from the operation count of the underlying detection algorithm divided by the reported computational throughput (TOPS). Power consumption was derived from the reported energy efficiency (TOPS/W), and throughput, energy efficiency, and energy-per-bit metrics were subsequently obtained from the normalized latency and power estimates.

For ASIC implementations, scaling was performed according to the architectural organization reported in each work. In general, area and power were assumed to scale proportionally to the dominant computational resources, while throughput and latency were adjusted according to the corresponding algorithmic complexity and iteration count.

For the ADMIN detector~\cite{shahabuddin_admm-based_2021}, normalization was derived from the two operating points reported by the authors ($16\times16$ and $32\times32$, both evaluated using 64-QAM modulation). Latency, power consumption, and silicon area were extrapolated to larger dimensions using linear regressions fitted to the reported measurements. The resulting scaling trends were subsequently projected to the target $64\times64$ benchmark. Because only two operating points are available, the normalization should be regarded as a first-order estimate. Nevertheless, the measured data suggest approximately linear growth of latency with problem dimension, together with increasing power and area requirements as larger detector instances are considered. Throughput, energy efficiency, and area efficiency were then derived from the projected latency, power, and area values. Technology scaling was not applied unless explicitly stated, ensuring that the normalized results remain representative of the originally reported implementation.

For the MPD detector of Tang \textit{et al.}~\cite{tang_058-mm2_2021}, the architecture comprises iterative processing elements and grouped layer-parallel computation blocks whose complexity scales approximately quadratically with the number of users. Area and power were therefore assumed to scale as $N_t^2$, while throughput scales linearly with the number of detected symbols and inversely with the number of iterations required for convergence. Energy-per-bit was derived from the resulting throughput and power estimates.

For the LAMA detector of Jeon \textit{et al.}~\cite{jeon_354_2019}, complexity is dominated by operations on the Gramian matrix, resulting in a quadratic dependence on the number of users. Area and power were again assumed to scale as $N_t^2$, whereas throughput scales proportionally to the number of detected bits and inversely with the iteration count.

For the mixed-signal IMC architecture of Zuo \textit{et al.}~\cite{zuo_precise_2025}, normalization was performed using the Cholesky-based complexity model reported in the supplementary information. Because the architecture is dominated by peripheral circuitry, area was assumed to scale approximately linearly with matrix dimension, consistent with the trends reported by the authors. Power scaling was similarly derived from the measured dependence on problem size. The reported implementation targets real-valued matrices; therefore, a factor of two was introduced when mapping complex-valued MIMO systems to their equivalent real-valued representation.

For the JETCAS'22 architecture~\cite{mannocci_analogue_2022}, latency was estimated from the operation count reported by the authors divided by the measured throughput, while power and area were reconstructed from the reported TOPS/W and TOPS/mm$^2$ figures. Since the architecture was originally evaluated on a $128\times64$ system, this operating point was used as the closest equivalent to the normalized $64\times64$ benchmark.

Whenever iteration counts were not explicitly reported for the target operating point, scaling was conservatively estimated from the corresponding loading ratio and detector characteristics. Because convergence behavior generally depends on channel statistics and signal-to-noise ratio, these normalized results should be interpreted as first-order projections rather than exact reproductions of the original hardware measurements. Nevertheless, the normalization procedure preserves the dominant complexity trends and enables a meaningful comparison of throughput, energy efficiency, area efficiency, and energy-per-bit across otherwise incompatible architectures.

%% file: supplementary.tex
\renewcommand{\thefigure}{\arabic{figure}}
\renewcommand{\figurename}{Extended Data Figure}
\renewcommand{\theHfigure}{ED\arabic{figure}}
\setcounter{figure}{0}

\begin{figure}
\centering
\includegraphics[width=\textwidth]{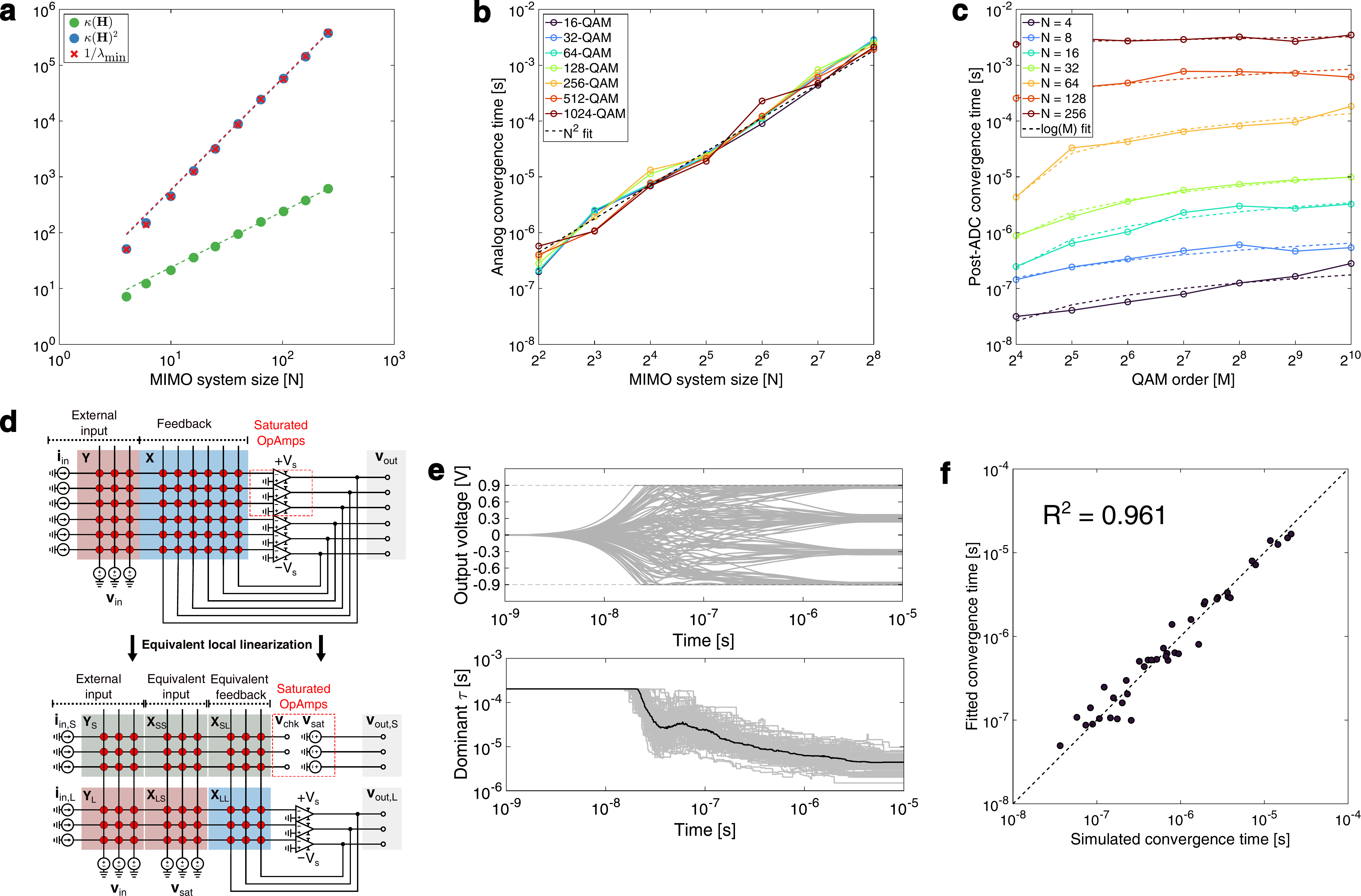}
\caption{\hl{\textbf{Convergence time analysis of BCZF circuit.} (a) Condition number $\kappa$ of Gaussian channel matrices $\mat{H}$, squared condition number $\kappa^2$, and reciprocal of the real part of the minimal eigenvalue as a function of MIMO system size, highlighting an $\mathcal{O}(N)$ dependence for $\kappa$ and corresponding $\mathcal{O}(N^2)$ dependence for $\kappa^2$ and $1/\lambda_{min}$. (b)~Analog convergence time of the equivalent fully-linear, supply-unbounded circuit as a function of MIMO system size. (c)~Post-ADC median convergence time of the equivalent fully-linear, supply-unbounded circuit as a function of QAM depth. (d)~Equivalent local linearization obtained upon saturation of one or more OpAmps, where only a submatrix remains active in a reduced linear feedback subsystem. (e)~Example transient for a 64×64, 16-QAM circuit (top), and corresponding dynamical transients for the dominant circuit time-constant, for 100 circuit runs with the same channel matrix and different input vectors (bottom; solid black line denotes the median). Saturation of one or more amplifiers causes a rearrangement of circuit poles, leading to transient acceleration over time. (f)~Convergence times obtained by Eq.~\eqref{eq:tts_fit} as a function of median convergence times obtained by circuit simulation.}}
\label{exfig:tts}
\end{figure}

\clearpage
\begin{figure}
\centering
\includegraphics[width=\textwidth]{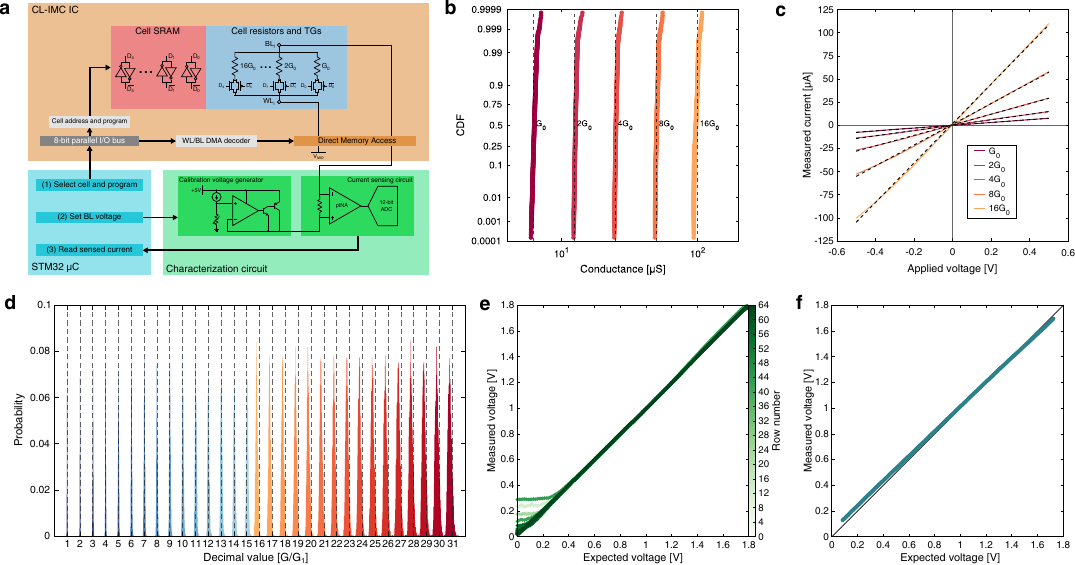}
\caption{\textbf{Experimental testchip characterization.} (a)~Board schematic, comprising the 90 nm experimental testchip for matrix-vector multiplication and inversion, dedicated circuitry for on-chip memory and peripherals characterization, and a microcontroller unit for board control and interfacing. (b)~Cumulative distribution functions and (c)~I-V characteristics of the integrated resistors employed in the 40T5R memory cells, highlighting low variability and high linearity. (d)~Histograms of the programmable conductance levels of the 40T5R memory cell, characterized on a total of 4096 cells. (e,f)~Conversion characteristics of on-chip (e)~DACs and (f)~ADCs.}
\label{exfig:chip}
\end{figure}

\clearpage
\begin{figure}
\centering
\includegraphics[width=\textwidth]{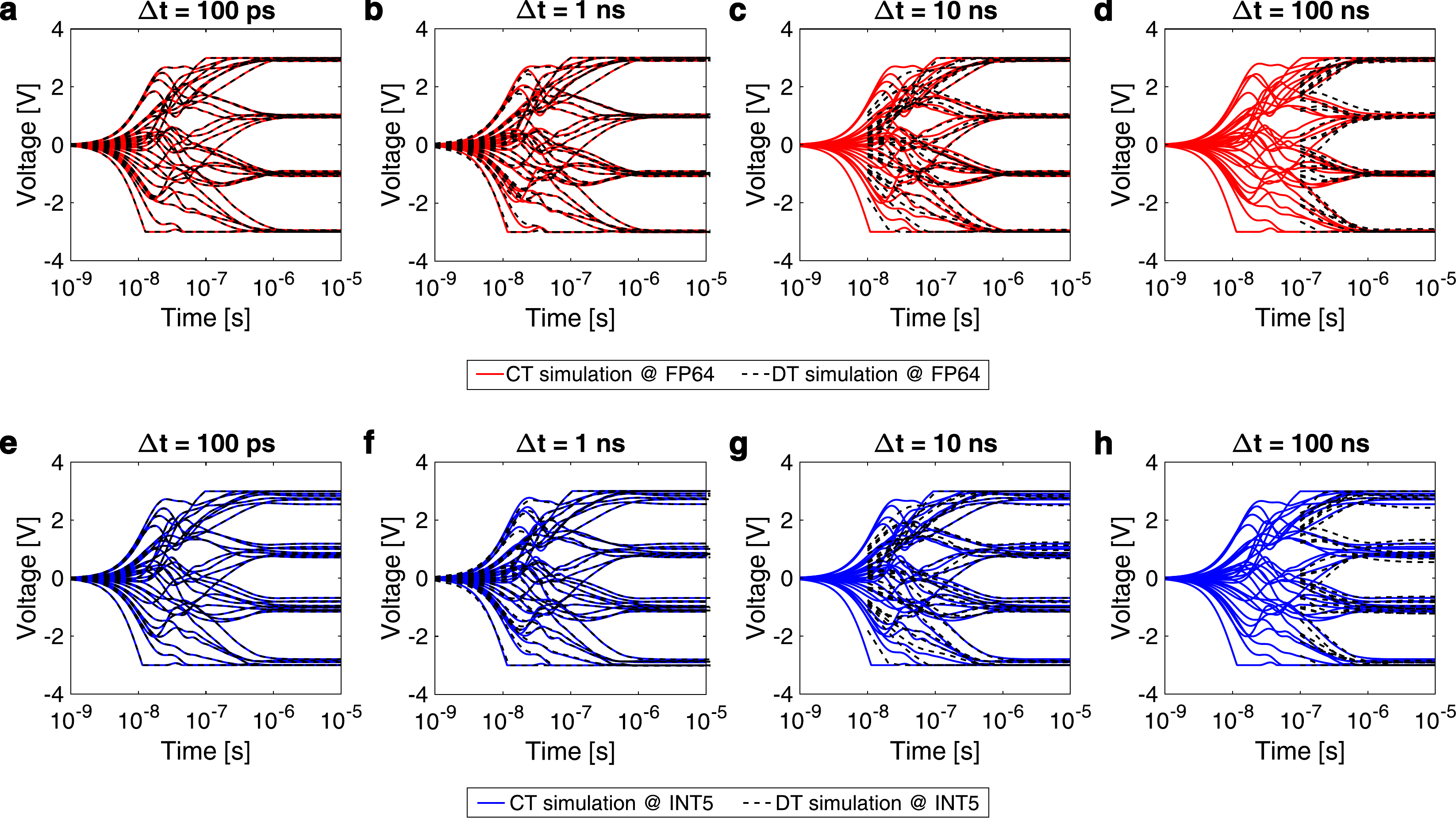}
\caption{\textbf{Continuous-time vs discrete-time model.} (a-d)~Output voltage transient predicted by the continuous-time model and by the two-step discrete-time model of Eq.~\eqref{eq:dt_2step} for a 16×16 16-QAM MIMO system at full FP64 channel precision and (e-h)~at reduced INT5 channel precision. For sufficiently small update steps ($\Delta t$~=~100~ps–1~ns), the discrete-time model accurately reproduces the continuous-time dynamics at both FP64 and INT5 precision. As $\Delta t$ increases, the trajectories progressively deviate owing to the operator-splitting approximation used to map the continuous-time dynamics onto experimentally measurable MVM and IMVM operations.}
\label{exfig:ctvsdt}
\end{figure}

\clearpage
\begin{figure}
\centering
\includegraphics[width=\textwidth]{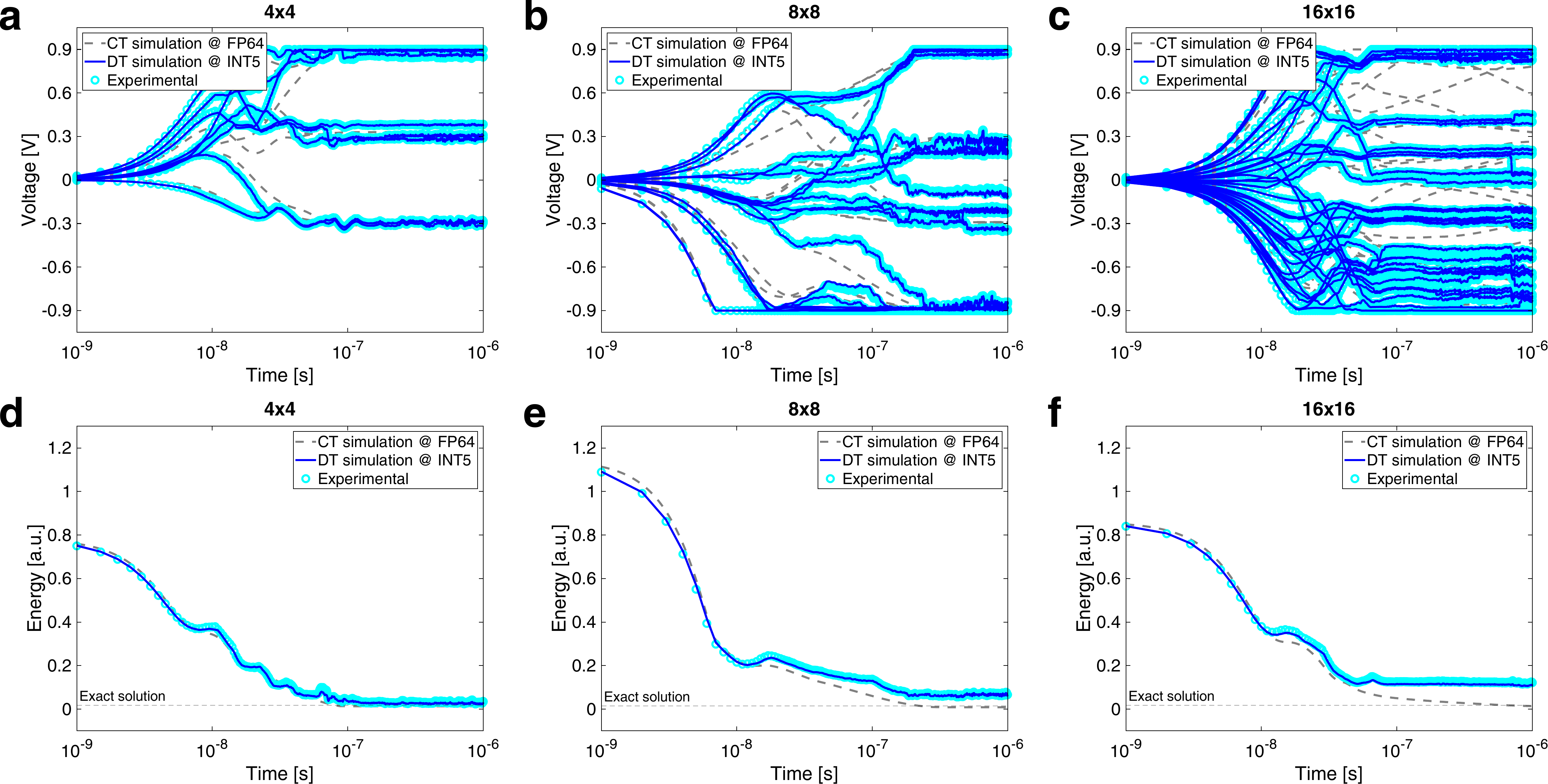}
\caption{\textbf{Experimental results at increasing channel size.} (a-c)~Comparison of output voltage transients and (d-f)~corresponding energy transient as computed by the CT-model at full FP64 precision (dashed grey), DT model (Eq.~\eqref{eq:dt_2step}) at IC-equivalent INT5 precision (solid blue), and experimental results obtained by on-chip solution (cyan).}
\label{exfig:expscale}
\end{figure}

\clearpage
\begin{figure}
\centering
\includegraphics[width=\textwidth]{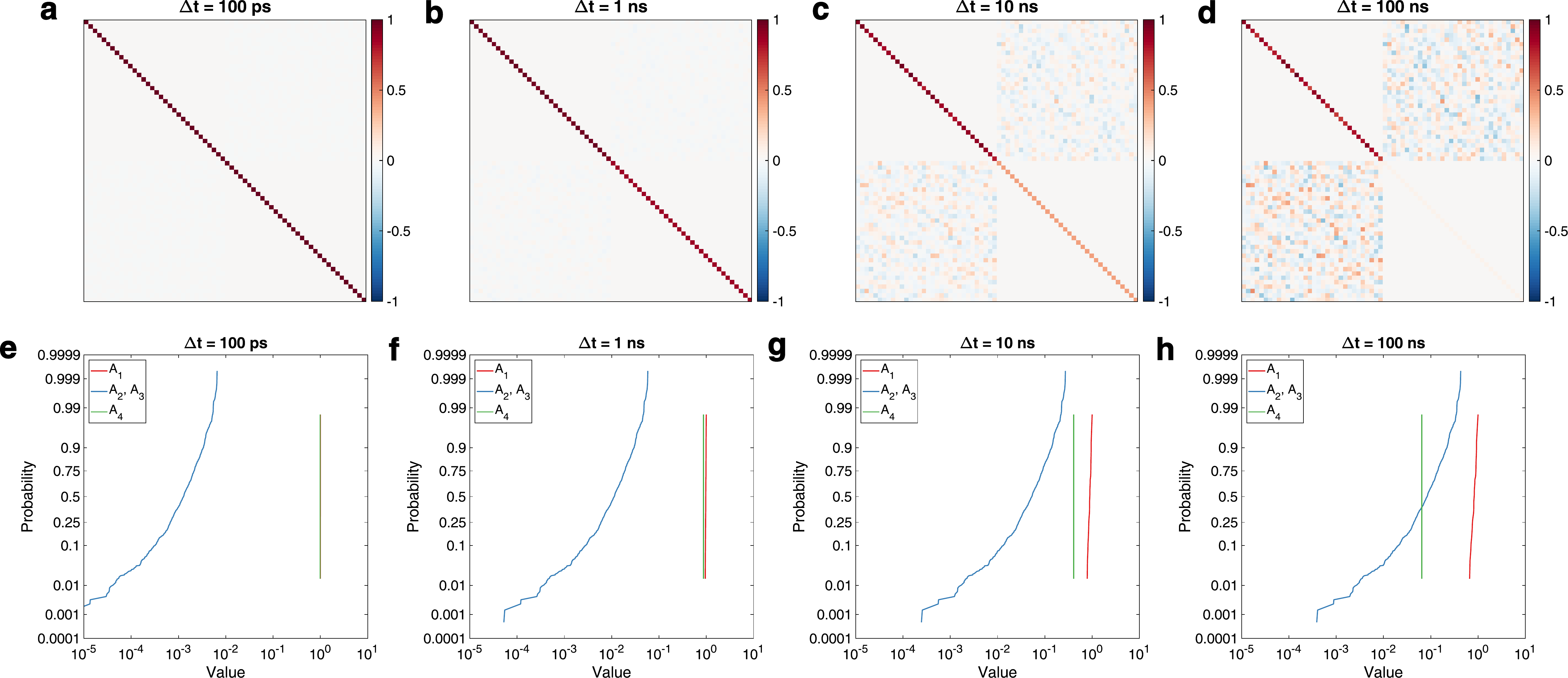}
\caption{\textbf{Numerical range of DT model matrix.} (a-d)~Step update matrices $\mat{A}$ of the DT model and (e-h)~cumulative distribution functions of the corresponding numerical values for sub-matrices $\mat{A}_1, \mat{A}_2, \mat{A}_3, \mat{A}_4$ of the update matrix $\mat{A}$ for different values of discretization step $\Delta t$, namely (a,e)~$\Delta t = $100~ps, (b,f)~$\Delta t = $1~ns, (c,g)~$\Delta t = $10~ns, (d,h)~$\Delta t = $100~ns.}
\label{exfig:dtmat}
\end{figure}

\clearpage
\begin{figure}
\centering
\includegraphics[width=\textwidth]{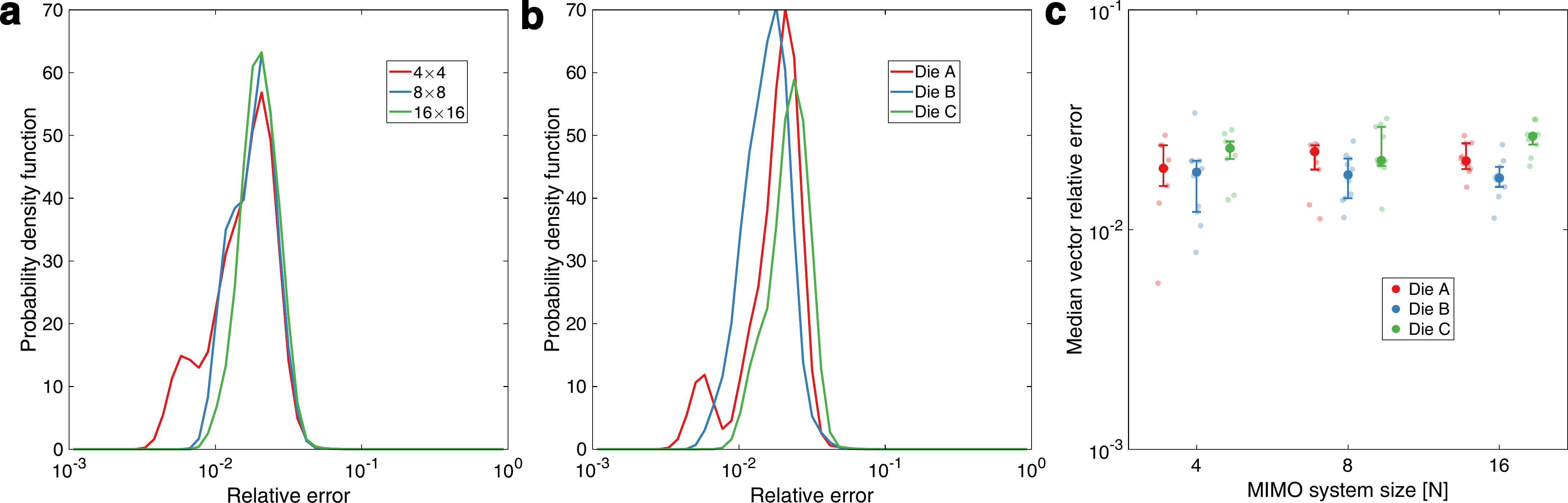}
\caption{\hl{\textbf{Experimental accuracy across MIMO dimensions and dies.} (a)~Probability-density distributions of the time-resolved vector relative error for $4\times4$, $8\times8$, and $16\times16$ MIMO systems, pooled across three dies. (b)~Corresponding distributions for dies A--C, pooled across the three MIMO dimensions. Curves in (a,b) show Gaussian-smoothed, probability-density-normalized histograms. (c)~Error statistics for each MIMO-dimension and die combination. Each translucent point represents the median time-resolved vector relative error from one experimental run; solid markers and error bars indicate the median and interquartile range across runs, respectively. Measurements were acquired using $\Delta t=1\,\mathrm{ns}$, with 10 randomly generated channel realizations and test vectors per condition. Dies A, B, and C correspond to three physically distinct test chips.}}
\label{exfig:expaccuracy}
\end{figure}